\documentclass[aps,prd,amsmath,floats,floatfix,twocolumn,superscriptaddress,nofootinbib,showpacs]{revtex4-1}

\usepackage{amssymb,amsmath,verbatim,mathtools,needspace,enumitem,etoolbox,graphicx,microtype,afterpage,bm,revsymb}
\usepackage{mathrsfs}

\usepackage[dvipsnames, usenames]{xcolor}
\definecolor{linkcolor}{rgb}{0.1,0.2,0.6}
\usepackage[unicode, colorlinks=true, linkcolor=linkcolor, citecolor=linkcolor, filecolor=linkcolor,urlcolor=linkcolor, pdfusetitle]{hyperref}

\usepackage[T1]{fontenc}
\usepackage[utf8]{inputenc}
\usepackage{tabularx, booktabs}
\usepackage[italicdiff]{physics}
\usepackage[caption=false]{subfig}
\usepackage[normalem]{ulem}

\graphicspath{{./Figures/}}

\begin{document}

\title{Living on the Edge of Effective Field Theory: \\ Near-Extremal Black Holes in Quadratic Gravity}

\author{Kelvin~Ka-Ho~Lam} 
\email{khlam4@illinois.edu}
\affiliation{Illinois Center for Advanced Studies of the Universe \& Department of Physics, University of Illinois Urbana-Champaign, Urbana, Illinois 61801, USA}

\author{Gary~T.~Horowitz} 
\email{horowitz@ucsb.edu}
\affiliation{Department of Physics, University of California, Santa Barbara, CA 93106, U.S.A.}

\author{Nicol\'as~Yunes}
\email{nyunes@illinois.edu}
\affiliation{Illinois Center for Advanced Studies of the Universe \& Department of Physics, University of Illinois Urbana-Champaign, Urbana, Illinois 61801, USA}

\date{\today}

\begin{abstract}

Near-extremal black holes can amplify higher-derivative corrections to general relativity, making them sharp probes of the perturbative control of gravitational effective field theory.
We study this question in quadratic gravity with a dynamical scalar degree of freedom, focusing on dynamical Chern-Simons and scalar Gauss-Bonnet gravity.
Using pseudospectral methods, we construct rapidly rotating exterior solutions at fixed surface gravity and fixed horizon angular velocity, and follow this controlled family toward extremality.
In dynamical Chern-Simons gravity, the sequence approaches an asymptotically flat extremal exterior whose near-horizon limit agrees with the dynamical-Chern-Simons-deformed near-horizon extremal Kerr geometry and inherits its enhanced $SO(2,1)\times U(1)$ symmetry.
In scalar Gauss-Bonnet gravity, the same limiting procedure behaves differently: the scalar develops an unavoidable logarithmic singularity at the horizon, the exterior metric develops a small boundary layer near the horizon, and the scalar Gauss-Bonnet-deformed near-horizon extremal solution is not connected to a regular asymptotically flat exterior.
We then use curvature invariants and tidal forces to diagnose the breakdown of the perturbative effective-field-theory expansion.
We find that scalar curvature invariants can remain finite at the horizon even when tidal forces measured by infalling observers diverge as inverse powers of the surface gravity. 
These results separate distinct notions of effective-field-theory breakdown near extremality and provide a perturbative validity estimate that can be compared with current bounds on the dynamical Chern-Simons and scalar Gauss-Bonnet couplings.

\end{abstract}

\maketitle

\section{Introduction}\label{sec:introduction}

General relativity (GR) has passed many high-precision tests in the low-curvature regime \cite{Will2014,Yunes:2013dva}, but the high-curvature regime remains much less explored.
This is precisely where corrections to the Einstein-Hilbert action are expected to become most important.
There are two complementary ways to organize such corrections.
From a top-down perspective, candidate ultraviolet (UV) completions, such as string theory, produce an infinite tower of higher-curvature operators in the low-energy action after compactification \cite{Gross_Sloan_1987,Gross:1986iv,Cano:2019ozf,Cano:2021rey,Maeda:2009uy,Moura:2006pz}.
From a bottom-up perspective, effective field theory (EFT) incorporates all local, symmetry-preserving operators allowed by diffeomorphism and Lorentz invariance, given the assumed field content \cite{Donoghue_2012}.
The resulting action is organized as a Wilsonian expansion in higher-curvature operators, characterized in geometric units by a short length scale $\ell$ associated with the UV physics.
When evaluated on a background with characteristic length scale $\lambda$, the perturbative expansion is controlled by powers of $\ell/\lambda$.
In this sense, the low-energy imprint of any UV completion that respects the symmetries of GR can be encoded through an appropriate choice of EFT couplings.

Recently, near-extremal black holes (BHs) have emerged as a useful arena in which to study the amplification of new physics in gravitational EFTs.
In some higher-derivative theories, tidal forces at the event horizon diverge as inverse powers of the surface gravity $\kappa$, even when scalar curvature invariants remain finite \cite{Horowitz:2023xyl,Horowitz:2024dch,Horowitz:2026fzc}.
In other cases, scalar curvature invariants themselves can become large near extremality, and EFT corrections to the quasinormal-mode (QNM) spectrum can become amplified.
In particular, the phase boundary between damped QNMs and zero-damped QNMs can shift \cite{Yang:2012he,Yang:2012pj}, producing parametrically enhanced QNM lifetimes \cite{Cano:2025ejw,DiRusso:2025qpf,Boyce:2025fpr,Husken:2026axq,vanderSteen:2026qva,Boyce:2026rnn}.
Some modes may also receive positive corrections to their imaginary parts that overwhelm the Kerr decay rate, leading to linear instabilities \cite{Husken:2026axq,vanderSteen:2026qva,Boyce:2025fpr,Boyce:2026rnn}.

These features are exciting because they can make beyond-GR effects observationally relevant, but they also raise questions about perturbative control.
Given a quantity $A$, the leading-order correction $(\ell/\lambda)^p A^{(1)}$ should remain smaller than the GR contribution $A^{(0)}$ \cite{Stein:2014xba,Cano:2025ejw,Chen:2024sgx}, where $p$ is a positive integer determined by the mass dimension of the higher-curvature operator.
This small-deformation, or small-coupling, approximation must be enforced when working with EFTs of gravity.
Neglecting this perturbative hierarchy can lead to misinterpretations, for example in causality ``constraints'' inferred from EFT corrections to Shapiro time delays \cite{Camanho:2014apa,Serra:2022pzl,Cassem:2026dCSBounds}.
If the truncated EFT is pushed beyond its weak-gravity, eikonal, or cutoff regime, apparent time advances need not be controlled observables of the EFT and can be mistaken for constraints on the theory.
This was shown explicitly for dynamical Chern-Simons (dCS) gravity in~\cite{Alexander:2025weakgravity}.
Near extremality, the amplification of higher-derivative corrections can violate the same perturbative-control criterion at fixed coupling.
Thus, near-extremal and extremal BHs do not merely enhance possible deviations from GR; they also probe the boundary of validity of the EFT expansion.

In past literature, EFT-corrected near-extremal and extremal BHs were studied by two different approaches. 
In the first method, one constructs the full exterior solution by solving the linearized field equations in a GR background. 
This approach has been used to study primarily Reissner-Nordstr\"{o}m BHs, with various boundary conditions -- asymptotically flat \cite{Yajima:2000kw, Brenna:2012gp, Cano:2020ezi, BarrosoVarela:2023ull} or anti-de Sitter \cite{Fernandes:2020rpa, Cremonini:2019wdk, Chen:2024sgx} -- in four and arbitrary spacetime dimensions. 
This is because the field equations are a set of coupled {\it ordinary} differential equations, where exact analytic solutions can often be found. 
On the other hand, rotating (charged or neutral) solutions in higher-derivative EFTs have only been studied recently by constructing numerical solutions \cite{Horowitz:2023xyl, Horowitz:2024dch}. 
Because of the reduced symmetry, from spherical symmetry to axisymmetry (henceforth we restrict to four dimensions), analytic metric solutions in the exterior spacetime do not exist or are difficult to obtain, and the extremal solution cannot be obtained straightforwardly. 
Note that the extremal limit of rotating BHs was studied in some modified gravity theories, but those theories are mostly treated as exact theories instead of EFTs \cite{Kleihaus:2011tg, Kleihaus:2015aje, Cano:2019ozf, Delsate:2018ome}. 

The second approach bypasses this difficulty by only computing the near-horizon extremal solution.
In GR, it is well known that the near-horizon region of an extremal BH can be described by the near-horizon extremal Kerr (NHEK) metric, which has an enhanced isometry $SO(2,1)\times U(1)$ \cite{Bardeen:1999px}. 
Extending to EFTs, one can construct EFT-deformed NHEK solutions, such that the same enhanced symmetry is still manifest \cite{Astefanesei:2006dd, Kunduri:2007vf, Kunduri:2013gce}. 
This approach was adopted in, e.g., \cite{Chen:2018jed, Cano:2024bhh, Cano:2023dyg, Cano:2019ozf, Horowitz:2023xyl, Horowitz:2024dch}. 
Due to the enhanced symmetry, the field equations depend only on the polar angle, leading to a set of ordinary differential equations that can be solved analytically \cite{Chen:2018jed, Horowitz:2023xyl, Horowitz:2024dch, Cano:2024bhh, Chen:2024sgx}. 
However, in general, these EFT-deformed NHEK solutions could be disconnected from an exterior spacetime with boundary conditions of astrophysical BHs, e.g., asymptotically flat and with no conical singularities \cite{Kunduri:2013gce,Cano:2019ozf,Horowitz:2024kcx}.
This is because the near-horizon problem is local: it fixes the throat geometry, but it does not guarantee that the modes needed to glue the throat to an exterior are compatible with the asymptotic boundary conditions.

In this work, we investigate EFT-corrected (near-)extremal BHs and the breakdown of EFT using both approaches. 
First, we construct the full exterior metric solutions numerically and subsequently analyze the near-horizon limit of the extremal geometries.
We focus on dCS gravity \cite{dCS_01, dCS_02} and scalar Gauss-Bonnet (sGB) gravity \cite{Yagi:2015oca, Nair:2019iur}. 
These theories are higher-derivative EFTs with dynamical pseudoscalar and scalar fields respectively, which are coupled to gravity through quadratic curvature invariants $\mathscr{Q} \sim {\rm Riem}^2$ \cite{Yunes:2011we, Cano_Ruiperez_2019}. 
These theories are well-motivated because they emerge as the low-energy limit of different quantum gravity theories, such as string theory \cite{Maeda:2009uy, Moura:2006pz, Cano_Ruiperez_2019, Cano:2021rey,Sen:2026xfy} or loop quantum gravity \cite{dCS_01, Taveras:2008yf}. 
From the EFT perspective, the dCS and sGB corrections are the leading-order corrections of the Einstein-Hilbert action that contains quadratic-in-Riemann terms with one dynamical scalar field \cite{Cano_Ruiperez_2019}. 

Previous work on rotating BHs in these theories followed two complementary routes.
One route constructed analytic solutions in a slow-rotation expansion, beginning with the slowly-rotating BHs of dCS and sGB gravity
\cite{Yunes:2009hc,Yagi:2012ya,Ayzenberg:2014aka,Maselli:2015tta,Cano_Ruiperez_2019}.
This route is powerful because it gives analytic control over the scalar field, the metric corrections, the conserved charges, and the small-coupling expansion.
The price is that the expansion parameter is the dimensionless spin $a=J/M^2$, where $M$ and $J$ are the Arnowitt-Deser-Misner (ADM) mass and angular momentum.
Although these series have now been pushed to very high order
\cite{Chung:2024ira,Chung:2024vaf,Chung:2025gyg}, comparisons with numerical solutions show that their practical accuracy deteriorates before the near-extremal regime is reached, at dimensionless spins of around $0.7$, even when terms beyond fortieth order in spin are included
\cite{Cano:2023qqm,Lam:2025fzi}.

A second route constructs the full rotating exterior numerically.
In previous work, two of us used spectral and pseudospectral methods to obtain accurate dCS and sGB BH solutions up to spins very close to unity
\cite{Lam:2025elw,Lam:2025fzi}.
Those earlier solutions reached very large spin, but they were not formulated to take the zero-surface-gravity, extremal limit.
In particular, the family of solutions was labeled by quantities defined at spatial infinity, while the extremal endpoint is more cleanly approached by fixing horizon data.
Here, we therefore modify the metric ansatz and the boundary conditions: instead of fixing the ADM mass and angular momentum, we fix the surface gravity and the horizon angular velocity.
As the surface gravity is decreased, the inner and outer horizons merge and the BH approaches extremality.

The first result of this paper is methodological.
Motivated by recent work showing that horizon data provide the natural way to approach extremality in higher-derivative BHs \cite{Horowitz:2023xyl,Horowitz:2024dch,Horowitz:2024kcx}, we introduce a horizon-adapted metric ansatz and boundary conditions that fix the surface gravity $\kappa$ and horizon angular velocity $\Omega_H$.
This allows us to construct controlled, asymptotically flat, rapidly rotating dCS and sGB exterior solutions much closer to extremality than previous implementations, reaching $\kappa \approx 10^{-4}$, or equivalently $1 - \bar{a} \sim O(10^{-8})$, where $\bar{a}$ is the dimensionless spin parameter used to label the Kerr background in our ansatz.
In dCS gravity, the same construction can then be continued all the way to the extremal endpoint.

The first physical result is that the dCS family has a continuous extremal endpoint, at least at leading order in the EFT expansion.
The scalar field and the metric corrections remain finite as $\kappa \to 0$, and the sub-extremal exterior solution connects to a directly constructed extremal solution.
Taking the near-horizon limit of this extremal exterior gives the dCS-deformed NHEK solution found previously in~\cite{Chen:2018jed,Cano:2023dyg}.
Thus, in dCS gravity, the EFT-corrected extremal BH is not merely a local throat geometry.
It exists as an asymptotically flat exterior spacetime whose throat is continuously connected to the deformed NHEK solution.

The sGB case is qualitatively different. 
At any finite, nonzero surface gravity, the solutions remain regular, but the limit $\kappa \to 0$ is singular. 
The scalar field develops an unavoidable logarithmic singularity at the horizon, while the subextremal exterior metric develops a small boundary layer of width $\kappa$ near the horizon, thus the subextremal metric corrections are not connected to the extremal solution. 
At extremality, the sGB-deformed NHEK scalar field obtained by solving the field equations directly in the near-horizon throat differs from the near-horizon limit of the full extremal exterior scalar field, as the two satisfy different boundary conditions. 
Because these extremal scalar field solutions do not coincide, the corresponding sGB-deformed NHEK metric cannot be interpreted as the near-horizon limit of an astrophysical black hole exterior in shift-symmetric sGB gravity.

Our numerical, near-extremal solutions also allow us to separate two notions of EFT breakdown.
In both theories, scalar curvature invariants remain controlled at the horizon as $\kappa \to 0$, but tidal forces measured by freely falling observers diverge as inverse powers of $\kappa$.
We use these curvature and tidal diagnostics to estimate the regime of validity of the perturbative EFT expansion, finding that the strict extremal BH solution lies outside perturbative control, while current observational bounds on the dCS and sGB couplings remain within the EFT regime for the BHs observed so far.

The remainder of this paper is organized as follows.
In Sec.~\ref{sec:EFT}, we review dCS and sGB gravity and define the perturbative expansions used throughout the paper.
In Sec.~\ref{sec:Spectral}, we describe the horizon-adapted metric ansatz, boundary conditions, pseudospectral construction, and convergence tests.
In Sec.~\ref{sec:dCS}, we present the dCS pseudoscalar field, metric corrections, conserved charges, extremal solution, and near-horizon limit.
In Sec.~\ref{sec:sGB}, we present the corresponding sGB scalar field and metric corrections, and explain why the exterior-connected near-horizon limit differs from the standard sGB-deformed NHEK branch.
In Sec.~\ref{sec:BreakdownEFT}, we classify the relevant notions of EFT breakdown and estimate the regime of validity using curvature and tidal diagnostics.
Appendix~\ref{sec:Lambda} gives details of the extremal ansatz and the near-horizon expansion of the sGB solution. Appendix~\ref{sec:sGB_Scalar_Field} derives the analytic solution of the full exterior extremal sGB scalar field, and the final appendix checks the tidal divergence using the norm of the tidal tensor.

In the remainder of the paper, we use the following conventions: $x^{\mu} = (t, r, \chi, \phi)$, where $\chi = \cos\theta$ and $\theta$ is the polar angle; the signature of the metric tensor $g_{\mu\nu}$ is $(-, +, +, +)$; geometric units are used where $G = 1 = c$. 
For any quantities $A$ that appear in this paper, we write the corresponding GR quantity as $A^{(0)}$ and the leading-order-in-$\zeta$ correction as $A^{(1)}$:
\begin{equation}
    A = A^{(0)} + \zeta A^{(1)} + O(\zeta^2), 
\end{equation}
where $\zeta$ is a coupling that will be defined in Eq.~\eqref{eq:LEE}.

\section{Effective Field Theory Extensions of General Relativity}\label{sec:EFT}

The Lagrangian density for shift-symmetric massless dCS and sGB gravity can be written as \cite{Lam:2025elw}
\begin{equation}\label{eq:Lagrangian}
    16 \pi \mathscr{L}_q = R - \frac{1}{2} (\nabla \varphi_q)^2 + \alpha_q \varphi_q \mathscr{Q}_q, 
\end{equation}
where $q = \text{dCS or sGB}$, and $\alpha_q = \ell_q^2$, where $\ell_q$ is the length scale of the EFT, $\varphi_q$ is a pseudoscalar or scalar field in dCS and sGB gravity, respectively, and $\mathscr{Q}_q$ are topological invariants,   
\begin{align}
    \mathscr{Q}_{\rm dCS} &= {}^*\!R_{\alpha\beta\gamma\delta}R^{\alpha\beta\gamma\delta}, \\ 
    \mathscr{Q}_{\rm sGB} &= R_{\alpha\beta\gamma\delta}R^{\alpha\beta\gamma\delta} - 4 R_{\alpha\beta}R^{\alpha\beta} + R^2, 
\end{align}
where ${}^*\!R_{\alpha\beta\gamma\delta} \equiv \frac{1}{2} \epsilon_{\alpha\beta\rho\sigma} R^{\rho\sigma}{}_{\gamma\delta}$ is the dual Riemann tensor, with the Levi-Civita tensor defined through 
\begin{equation}
\epsilon^{\rho \sigma \alpha \beta} = \frac{1}{\sqrt{-g}} [\rho \sigma \alpha \beta], 
\end{equation}
where $g$ is the determinant of $g_{\mu \nu}$, and $[\rho \sigma \alpha \beta]$ is the totally antisymmetric Levi-Civita symbol, with convention $[t r \chi \phi] = 1$~\cite{Chung:2025gyg, Lam:2025fzi}. 
Varying the action with respect to the metric and fields, we obtain the following equations of motion:
\begin{align}
    \Box \vartheta_q + \mathscr{Q}_q &= 0, \label{eq:SFE} \\
    R_{\mu}{}^{\nu} + \zeta_q (\mathscr{S}_q)_{\mu}{}^{\nu} &= 0, \label{eq:LEE}
\end{align}
where $\vartheta_q = \varphi_q/\alpha_q$ is a rescaled scalar field, $\zeta_q = \alpha_q^2/\lambda^4 = (\ell_q/\lambda)^4$ is the dimensionless small-coupling parameter, and $\Box = \nabla_{\mu}\nabla^{\mu}$ is the d'Alembertian operator.
Here $\lambda$ is the characteristic length scale of the background. For the BH solutions below, we quote dimensionless couplings using the ADM mass as this reference scale.

The source tensor $(\mathscr{S}_q)_{\mu}{}^{\nu}$ contains the higher-derivative and scalar field corrections to the field equations \cite{Cano_Ruiperez_2019, Lam:2025fzi}. 
\begin{equation}
\begin{split}
    \lambda^{-4} (\mathscr{S}_{\rm dCS})^{\mu\nu} &= -4\Big[\left(\nabla_\sigma \vartheta_{\rm dCS} \right) \epsilon^{\sigma \delta \alpha(\mu|} \nabla_\alpha R^{|\nu)}{}_{\delta} \\
    &+ \left(\nabla_\sigma \nabla_\delta \vartheta_{\rm dCS} \right) {}^*\!R^{\delta (\mu \nu) \sigma}\Big] - \frac{1}{2}\nabla^{\mu} \vartheta_{\rm dCS} \nabla^{\nu} \vartheta_{\rm dCS} \\
    \lambda^{-4} (\mathscr{S}_{\rm sGB})_{\mu}{}^{\nu} &= \left[\delta_{\mu \lambda \gamma \delta}^{\nu \sigma \alpha \beta} - \frac{1}{2} \delta_{\mu}{}^{\nu}\delta_{\eta \lambda \gamma \delta}^{\eta \sigma \alpha \beta}\right] R^{\gamma \delta}{}_{\alpha \beta} \nabla^{\lambda} \nabla_{\sigma} \vartheta_{\rm sGB} \\
    &- \frac{1}{2}\nabla_{\mu} \vartheta_{\rm sGB} \nabla^{\nu} \vartheta_{\rm sGB}, 
\end{split}
\end{equation}
where $\delta^{\nu \sigma \alpha \beta}_{\mu \lambda \gamma \delta}$ is the generalized Kronecker delta, defined as 
\begin{equation}
\delta_{\mu_1 \mu_2 \mu_3 \mu_4}^{\nu_1 \nu_2 \nu_3 \nu_4} = \det 
\begin{pmatrix}
\delta_{\mu_1}^{\nu_1} & \delta_{\mu_2}^{\nu_1} & \delta_{\mu_3}^{\nu_1} & \delta_{\mu_4}^{\nu_1} \\[1mm]
\delta_{\mu_1}^{\nu_2} & \delta_{\mu_2}^{\nu_2} & \delta_{\mu_3}^{\nu_2} & \delta_{\mu_4}^{\nu_2} \\[1mm]
\delta_{\mu_1}^{\nu_3} & \delta_{\mu_2}^{\nu_3} & \delta_{\mu_3}^{\nu_3} & \delta_{\mu_4}^{\nu_3} \\[1mm]
\delta_{\mu_1}^{\nu_4} & \delta_{\mu_2}^{\nu_4} & \delta_{\mu_3}^{\nu_4} & \delta_{\mu_4}^{\nu_4} 
\end{pmatrix}. 
\end{equation}
Since the Lagrangian density is parity-even, and ${}^*\!R_{\alpha\beta\gamma\delta}$ is odd, $\vartheta_{\rm dCS}$ is a parity-odd pseudoscalar field; meanwhile $\vartheta_{\rm sGB}$ is a parity-even scalar field. 
However, the source tensors in both theories are parity-even. 

We solve this set of equations by expanding the metric and scalar field in powers of $\zeta$ (we suppress the subscript $q$ from now on), 
\begin{align}
    g_{\mu\nu} &= g_{\mu\nu}^{(0)} + \zeta g_{\mu\nu}^{(1)} + O(\zeta^2), \\
    \vartheta &= \vartheta^{(0)} + O(\zeta), 
\end{align}
where $g_{\mu \nu}^{(0)}$ is taken to be the Kerr metric in Boyer-Lindquist coordinates:
\begin{align}\label{eq:Kerr_metric}
    g_{\mu \nu}^{(0)} dx^{\mu} dx^{\nu} &= -\frac{\Sigma \Delta}{\cal A} dt^2 + \Sigma \left(\frac{1}{\Delta} dr^2 + \frac{1}{1 - \chi^2} d\chi^2\right) \nonumber \\
    &+ \frac{\cal A}{\Sigma} (1 - \chi^2) (d\phi - \omega dt)^2, 
\end{align}
where
\begin{align}
{\cal A} &= (r^2 + \bar{M}^2 \bar{a}^2)^2 - \bar{M}^2 \bar{a}^2 \Delta (1 - \chi^2), \\
\Sigma &= r^2 + \bar{M}^2 \bar{a}^2 \chi^2, \\
\Delta &= r^2 - 2\bar{M}r + \bar{M}^2 \bar{a}^2, \\
\omega &= 2\bar{M}^2\bar{a}r/{\cal A}.
\end{align}
Here $\bar{M}$ and $\bar{a}$ label the zeroth-order Kerr background.
The corresponding inner and outer horizons are
$r_{\pm}=\bar{M}(1\pm\sqrt{1-\bar{a}^2})$.
We reserve $M$, $J$, and $a=J/M^2$ for the ADM mass, ADM angular momentum, and dimensionless spin of the full EFT-corrected spacetime.
These quantities differ from $\bar{M}$ and $\bar{a}$ at $O(\zeta)$, as discussed below.

Linearizing Eqs.~\eqref{eq:SFE} and \eqref{eq:LEE}, we have a set of linear coupled partial differential equations of the form
\begin{align}
    E_{\vartheta} &\equiv \Box^{(0)} \vartheta + \mathscr{Q}^{(0)} = 0. \label{eq:SF0} \\
    E_{\mu}{}^{\nu} &\equiv [R_{\mu}{}^{\nu}]^{(1)} + [\mathscr{S}_{\mu}{}^{\nu}]^{(0)} = 0, \label{eq:EE1}
\end{align}
where zeroth-order quantities are evaluated with respect to the Kerr metric in Eq.~\eqref{eq:Kerr_metric}. 
We have also dropped the superscript on $\vartheta$ since we will only work up to zeroth order in the scalar field. 

To compute the leading-order in $\zeta$ correction to the metric, we parametrize the metric as follows: 
\begin{equation}\label{eq:metric}
\begin{split}
g_{\mu \nu} dx^{\mu} dx^{\nu} &= -\frac{\Sigma \Delta}{\cal A} [1 + \zeta H_1(r, \chi)] dt^2 \\
&+ [1 + \zeta H_3(r, \chi)] \Sigma \left(\frac{1}{\Delta} dr^2 + \frac{\Xi}{1 - \chi^2} d\chi^2\right) \\
&+ \frac{\cal A}{\Sigma} (1 - \chi^2) \Xi [1 + \zeta H_4(r, \chi)] \\
&\qquad \times \left[d\phi - \omega (1 + \zeta H_2(r, \chi)) dt\right]^2, 
\end{split}
\end{equation}
where $H_i(r, \chi)$ are EFT-corrections of the Kerr solution and we include 
\begin{equation}
    \Xi(r) = 1 + \zeta \frac{\Lambda \bar{M}^4}{r^4}, 
\end{equation}
in the metric.
We choose  
\begin{equation}
    \Lambda_{\rm dCS} = -\frac{975}{16\sqrt{2}}, \quad \Lambda_{\rm sGB} = -\frac{1097}{16\sqrt{2}}. 
\end{equation}
The value of $\Lambda$'s, derived in Appendix~\ref{sec:Lambda}, was found by requiring extremal BHs to have regular poles \cite{Horowitz:2024dch}. 

For an elliptic partial differential equation, we must impose suitable boundary conditions.
At the polar axis $\chi = \pm 1$, to ensure the absence of conical singularities, we require 
\begin{equation}\label{eq:AxisBC}
    H_3(r, \chi = \pm 1) = H_4(r, \chi = \pm 1).
\end{equation}
At spatial infinity, we define the following series expansion:
\begin{align} \label{eq:H_Inf_Expansion}
H_i(r, \chi) = {}_{(0)}H_i(\chi) + \frac{\bar{M}}{r} \; {}_{(1)}H_i(\chi) + O((r/\bar{M})^{-2})\,.
\end{align}
Then, asymptotic flatness and an asymptotic $S^2$ symmetry require, 
\begin{equation}\label{eq:InfinityBC}
    {}_{(0)}H_1 = 0, \quad {\rm and} \quad {}_{(0)}H_3 = {}_{(0)}H_4, 
\end{equation}
respectively.
Unlike in \cite{Lam:2025fzi}, where two of us have imposed another two boundary conditions at spatial infinity that fix the ADM mass and angular momentum, we here fix the surface gravity $\kappa$ and the angular velocity $\Omega_H$ at the horizon to be the same as those for a Kerr BH. Explicitly, 
we set 
\begin{align}\label{eq:HorizonBC}
    H_1(r_+, \chi) - H_3(r_+, \chi) = 0, \quad H_2(r_+, \chi) = 0, 
\end{align}
such that 
\begin{align}
    \kappa &= \kappa^{(0)} \left[1 + \frac{\zeta}{2}\big(H_1(r_+, \chi) - H_3(r_+, \chi)\big)\right] = \kappa^{(0)}, \\
    \Omega_H &= \Omega_H^{(0)} \left[1 + \zeta H_2(r_+, \chi)\right] = \Omega_H^{(0)}, 
\end{align}
where $\kappa^{(0)}=(r_+-\bar{M})/(2\bar{M}r_+)$ and $\Omega_H^{(0)}=\bar{a}/(2r_+)$ are the Kerr surface gravity and horizon angular velocity.

The solutions constructed this way are parametrized by $(\bar{M},\bar{a})$, or equivalently by $(\kappa,\Omega_H)$.
As $\bar{a}\to1$, the surface gravity $\kappa$ tends to zero, so this parametrization gives direct access to extremality.
By contrast, the previous formalism of~\cite{Lam:2025elw,Lam:2025fzi} fixed ADM quantities.
Those variables are natural for generic subextremal BHs, but they are not the most natural coordinates on the solution space near extremality, where higher-derivative corrections shift the extremality condition of $a = 1$ \cite{Cano:2024bhh,Kleihaus:2011tg,Kleihaus:2015aje}.
Moreover, approaching extremality at fixed $M$ and $J$ can produce spurious divergences in conserved charges \cite{Reall:2019sah}.
We therefore work at fixed $\kappa$ and $\Omega_H$.

The set of boundary conditions in Eqs.~\eqref{eq:AxisBC}, \eqref{eq:InfinityBC} and \eqref{eq:HorizonBC}, together with Eq.~\eqref{eq:EE1} and the scalar field solution, allow us to fully determine the metric corrections $H_i(r, \chi)$ in the exterior spacetime.

\section{Constructing Sub-extremal Black-Hole Solutions in Quadratic Gravity} \label{sec:Spectral}

In this section, we review how to construct BH solutions in quadratic gravity theories using pseudospectral methods \cite{Dias:2015nua, Fernandes:2022gde, Fernandes:2025vxg}, following mostly \cite{Lam:2025fzi,Lam:2025elw}. 
Furthermore, we modify and extend this method to construct BH solutions with fixed $\kappa$ and $\Omega_H$. 

\subsection{Scalar Fields}\label{subsec:Scalar_Field}

First, before finding the metric corrections, we must solve for the scalar field $\vartheta(r, \chi)$. 
Following the pseudospectral method detailed in \cite{Stein:2014xba, Lam:2025fzi}, we construct dCS and sGB scalar field solutions. 
Since we follow the same approach as \cite{Lam:2025fzi}, we shall not repeat the full details here. 
Importantly, the scalar field can be represented as 
\begin{equation}\label{eq:Scalar_Field_Ansatz}
    \vartheta(r, \chi) = \sum_{n = 0}^{N_z} \sum_{l = 0}^{N_\chi} c_{nl} r^{-l-1} T_n(z) P_l(\chi), 
\end{equation}
where $T_n(\cdot)$ and $P_l(\cdot)$ are Chebyshev and Legendre polynomials, and $z = {2r_+}/{r} - 1$ compactifies the radial coordinate by mapping $r = r_+$ and $r = \infty$ to $z = 1$ and $z = -1$ respectively. The constants
$N_z$ and $N_{\chi}$ are the spectral orders of the scalar field solution. 
Because of the parity of $\vartheta$, only the odd $P_l$ enter $\vartheta_{\rm dCS}$, while the even $l$ ones enter $\vartheta_{\rm sGB}$. 
The scalar field solutions constructed thus have been shown to converge exponentially and are accurate for any spins below $0.9999$ \cite{Lam:2025fzi, McNees:2015srl}. 
We show later in this section that the scalar field in near-extremal spacetime is still accurate and that exponential convergence remains. 

\subsection{Metric Corrections}

With the scalar field in hand, we can now compute the metric corrections by solving Eq.~\eqref{eq:LEE}. 
We assume $H_i(r, \chi)$ can be represented by a spectral series, 
\begin{equation}\label{eq:SpectralSeries}
    H_i(r, \chi) = \sum_{n=0}^{{\cal N}_z} \sum_{l = 0}^{{\cal N}_\chi} v_{nl}^i T_n(z) P_{2l}(\chi), 
\end{equation}
where ${\cal N}_z$ and ${\cal N}_{\chi}$ are the spectral orders of the solution. 
% \KL{Maybe we do not need the paragraph below for conciseness. }\\
The spectral expansion together with the scalar field solution, when substituted into Eq.~\eqref{eq:EE1}, allow us to schematically write the linearized field equation as 
\begin{equation}\label{eq:EE1H}
\begin{split}
    \sum_{i=1}^{4} \sum_{\alpha,\beta,\delta,\sigma} {\cal G}_{i,\delta,\sigma,\alpha,\beta}^{j} z^{\delta} \chi^{\sigma} \partial_{z}^{\alpha} \partial_{\chi}^{\beta} H_i(r, \chi)
    % \left[\sum_{n=0}^{{\cal N}_z} \sum_{l=0}^{{\cal N}_\chi} v_{n\ell}^{i} T_{n}(z) P_{2l}(\chi)\right]
    &= \sum_{\delta,\sigma} {\cal S}_{\delta,\sigma}^{j} z^{\delta} \chi^{\sigma}, 
\end{split}
\end{equation}
where ${\cal G}_{i,\delta,\sigma,\alpha,\beta}^j$ are coefficients that depend on $\bar{M}$ and $\bar{a}$, while ${\cal S}_{\delta,\sigma}$ are coefficients that depend on $\bar{M}$, $\bar{a}$, and $\Lambda$ and $c_{nl}$. 
Out of the 10 independent equations in Eq.~\eqref{eq:EE1}, 4 are satisfied trivially thanks to the form of the metric ansatz. The remaining 6 equations -- the $(t,t), (t,\phi), (\phi\phi), (r,r), (\chi,\chi), (r, \chi)$ components -- are solved simultaneously. These components are indexed by $j = 1, 2, ...,6$ respectively. 

Instead of projecting Eq.~\eqref{eq:EE1H} onto orthogonal basis functions, as demonstrated in \cite{Lam:2025elw,Lam:2025fzi}, we adopt the pseudospectral collocation method. 
Pseudospectral methods, although often presenting a slower convergence rate than spectral expansions, have the advantage of higher computational efficiency and are more versatile for imposing boundary conditions. 
Substituting Eq.~\eqref{eq:SpectralSeries} into Eq.~\eqref{eq:EE1H}, and evaluating them on the Chebyshev collocation grid \cite{boyd2013chebyshev}, 
\begin{equation}
\begin{split}
    z_r &= -\cos\left(\frac{r \pi}{{\cal N}_z}\right)\hspace{2mm}\quad r = 0, 1, \ldots, {\cal N}_z, \\ 
    \chi_s &= -\cos\left(\frac{s \pi}{2{\cal N}_\chi}\right) \quad s = 0, 1, \ldots, {\cal N}_{\chi},
\end{split}
\end{equation}
we obtain a set of $6({\cal N}_z + 1)({\cal N}_\chi + 1)$ equations with $4({\cal N}_z + 1)({\cal N}_\chi + 1)$ unknowns $v_{nl}^i$. 
We only have to evaluate on the grid with $\chi_s \in [-1, 0]$ since the spectral series only contains even Legendre polynomials. 
This eliminates the need to evaluate on points with $\chi_s > 0$, thus reducing the size of the problem in half. 

Since the equations are linear in $v_{nl}^i$, they can be written compactly in matrix form,
\begin{equation}
    \mathbb{D}_{JI} \mathbf{v}_I = \mathbf{s}_J,
\end{equation}
where $\mathbb{D}$ is a rectangular matrix containing the coefficients of $v_{nl}^i$ on the left-hand side of Eq.~\eqref{eq:EE1H} after substituting Eq.~\eqref{eq:SpectralSeries} into the field equations, $\mathbf{v}$ is the vector of spectral coefficients, and $\mathbf{s}$ is the source vector on the right-hand side.
The composite indices $I=(i,n,l)$ and $J=(j,p,s)$ enumerate the unknowns and the collocation equations, respectively.

To impose boundary conditions, we apply the so-called boundary-bordering technique \cite{boyd2013chebyshev}. 
At the border of the Chebyshev grid, we replace the field equations with boundary conditions. 
For example, at spatial infinity $z = -1$, rather than evaluating Eq.~\eqref{eq:EE1H}, we take one of the boundary conditions from Eq.~\eqref{eq:InfinityBC} and evaluate at $z = -1$. 
Doing this for all the boundary conditions, we obtain a new matrix equation 
\begin{equation}\label{eq:SpectralMatrixEq}
    \tilde{\mathbb{D}} \mathbf{v} = \tilde{\mathbf{s}}.
\end{equation}
The matrix $\tilde{\mathbb{D}}$ is also normalized, i.e., $\max_{I} \tilde{\mathbb{D}}_{JI} = 1$ for every row $J$, by rescaling each matrix row and $\tilde{\bf s}$. 
We have checked that the final boundary-bordered system has a trivial kernel, $\ker\tilde{\mathbb{D}}=\{0\}$, so that the rectangular matrix equation can be readily solved by \texttt{Mathematica}'s built-in function \texttt{LeastSquares}.  Without loss of generality, we fix $\bar{M} = 1$ for all numerical solutions and in the remaining article. 

\subsection{Numerical Implementations and Convergence}
\subsubsection{Implementation}

To compute the scalar field, we truncate the scalar field at spectral order $N_z = N_{\chi} = 50$ for spins $\bar{a} \leq 0.9999$, and at $N_z = N_{\chi} = 100$ for $0.9999 < \bar{a} < 1$ in dCS gravity.
In sGB gravity, because the scalar field magnitude becomes large at the horizon, more spectral basis functions along the radial direction are needed. 
We set $N_z = 300$ and $N_{\chi} = 100$ for $1 - 10^{-6} < \bar{a} < 1$. 
With these choices, the scalar-field residual defined below remains below $10^{-7}$ for all solutions used in the metric solve.
This ensures that errors in the scalar field do not dominate the residuals of the metric equations.
All other numerical settings are identical to those used in~\cite{Lam:2025fzi}.

Similarly, for the spectral expansion of the metric corrections, we truncate the angular sum at ${\cal N}_{\chi} = 16$ for $\bar{a} \leq 0.999$, and ${\cal N}_{\chi} = 32$ for $0.999 < \bar{a} < 1$. In dCS gravity, along the radial direction, ${\cal N}_z = 30$ for $\bar{a} \leq 0.99$, ${\cal N}_z = 60$ for $\bar{a} \leq 0.9999$, ${\cal N}_z = 100$ for $\bar{a} \leq 0.999995$, and increases to ${\cal N}_z = 180$ for the largest spin $\bar{a} = 1 - 3 \times 10^{-8}$. 
In sGB gravity, a higher spectral order is needed for near-extremal solutions. For $\bar{a} \leq 0.9999$, we set ${\cal N}_z$ to be the same as in dCS gravity. 
But as $\bar{a}$ approaches unity, ${\cal N}_z$ increases to $220$ for the largest spin we compute, $\bar{a} = 1 - 10^{-7}$. 
Due to limited computational resources, we refrain from computing solutions with any higher spin. 

\subsubsection{Convergence}

\begin{figure*}[t!]
    \centering
    \includegraphics[width=0.99\columnwidth]{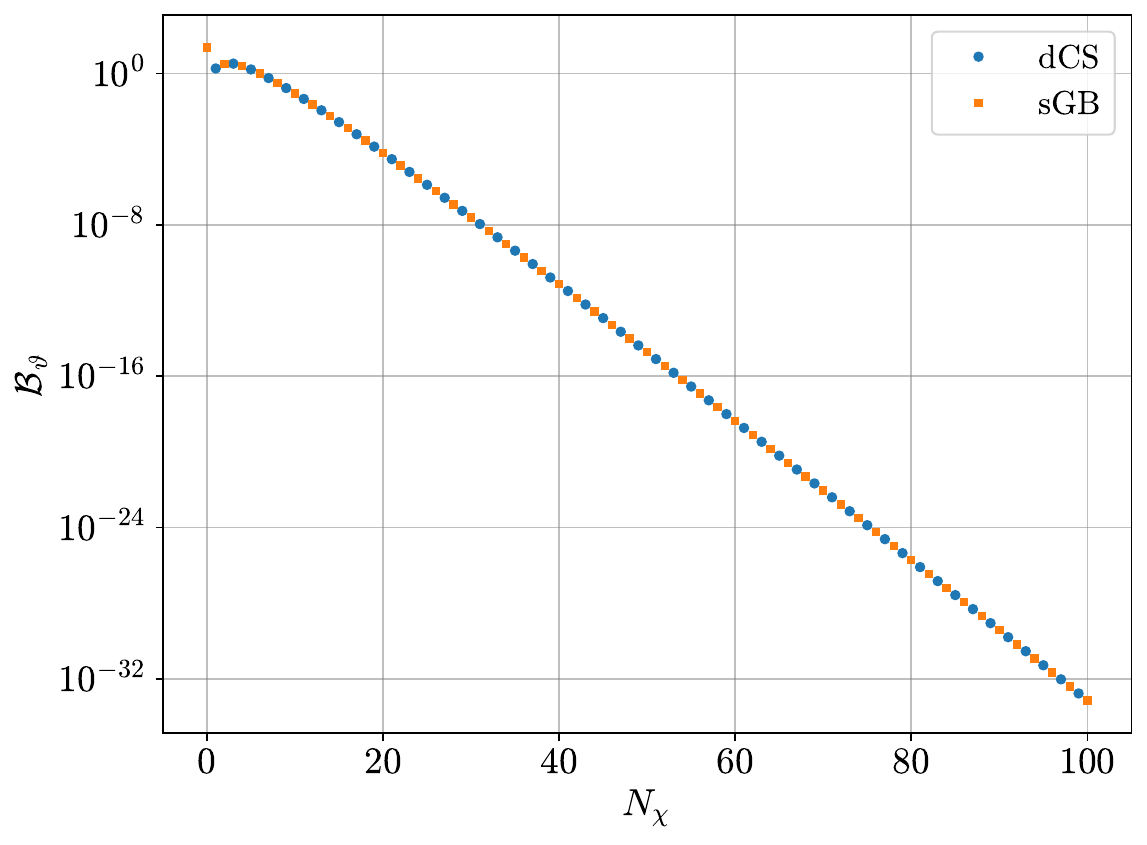}
    \hfill
    \includegraphics[width=0.99\columnwidth]{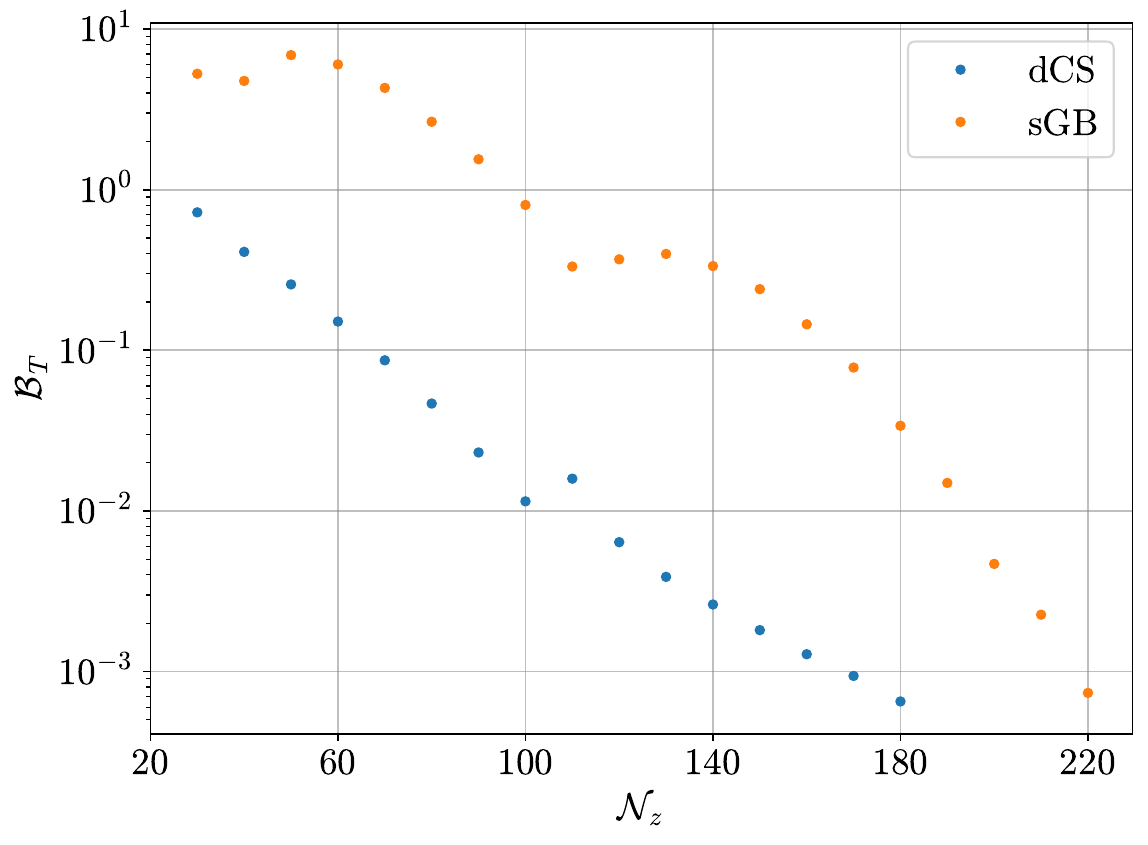}
    \caption{Backward modulus difference of dCS solutions with spin $\bar{a} = 1 - 3 \times 10^{-8}$, and sGB solutions with $\bar{a} = 1 - 10^{-7}$. Left: ${\cal B}_{\vartheta}$ against $N_{\chi}$, with fixed $N_z = 100$ and $300$, in dCS and sGB gravity respectively. Right: ${\cal B}_T$ against ${\cal N}_{z}$ with fixed ${\cal N}_{\chi} = 32$ in both theories. }
    \label{fig:BMD}
\end{figure*}

\begin{figure*}[t!]
    \centering
    \includegraphics[width=0.99\columnwidth]{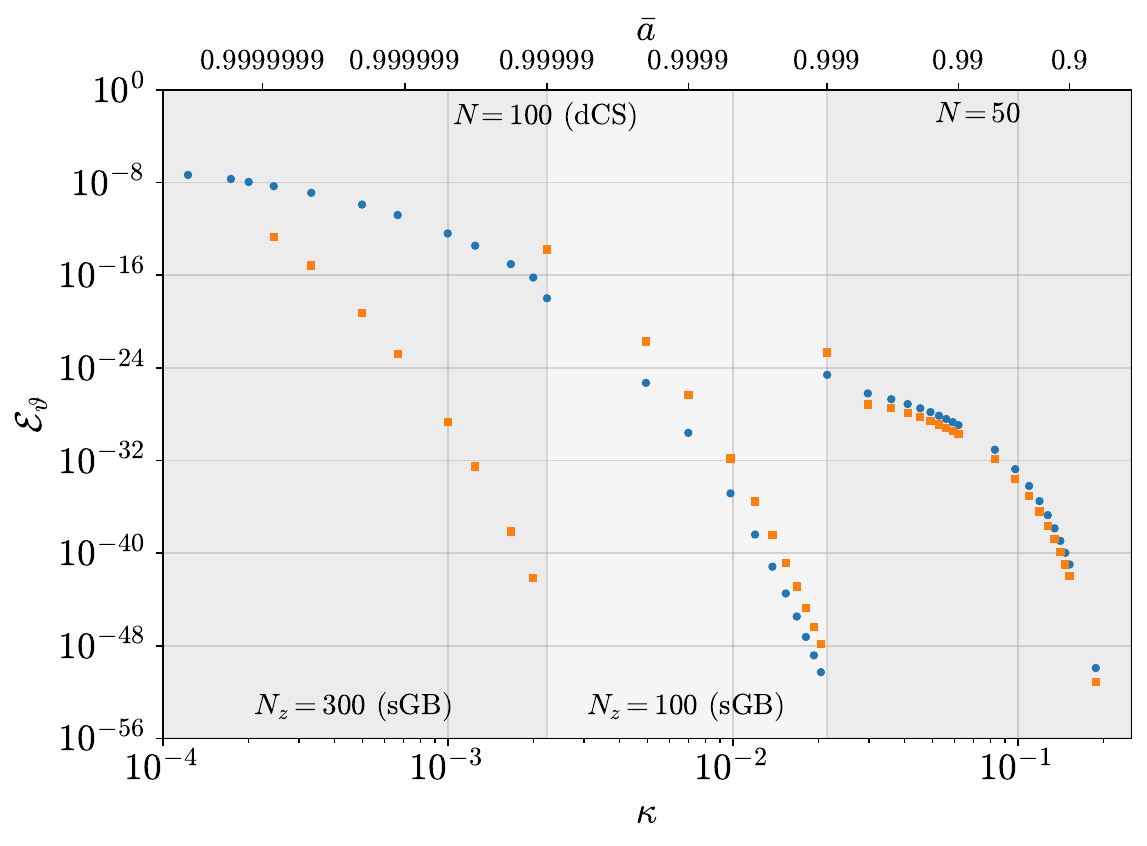}
    \hfill
    \includegraphics[width=0.99\columnwidth]{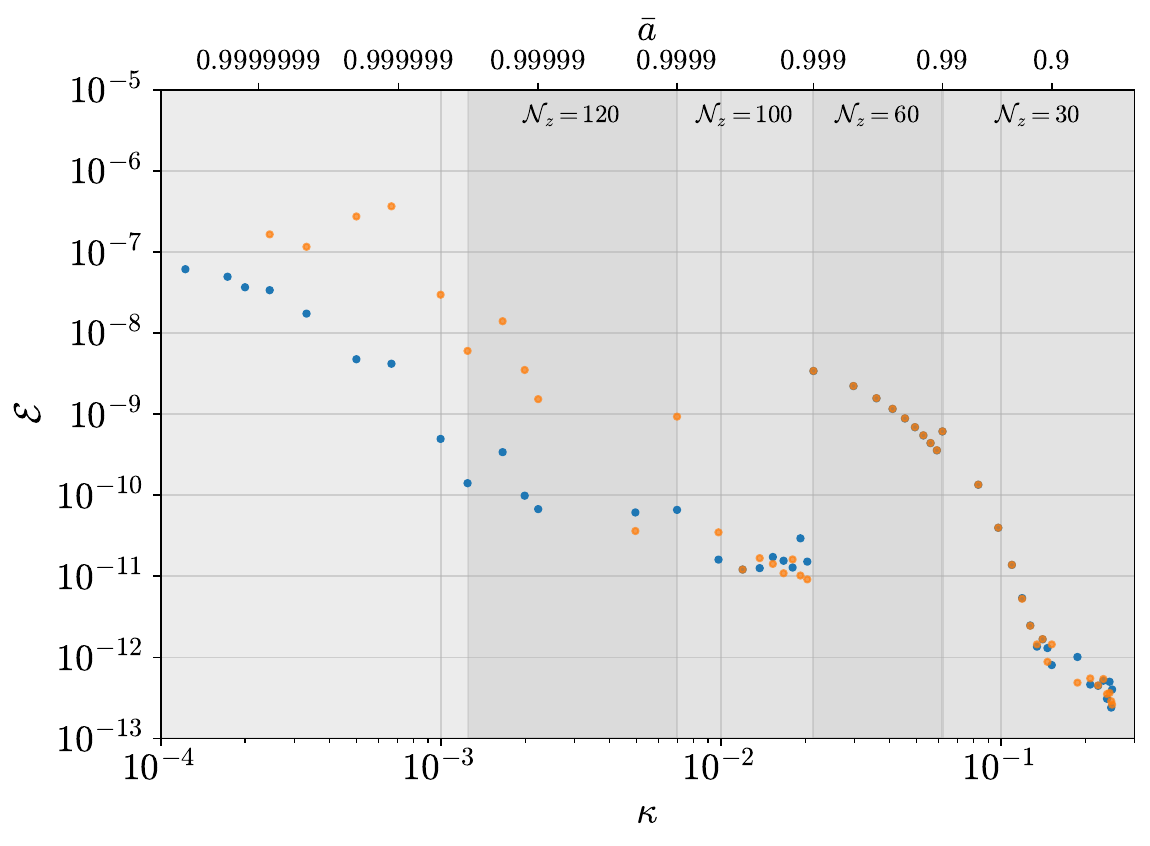}
    \caption{Error residual of the scalar field equation ${\cal E}_{\vartheta}$ (left) and the field equation ${\cal E}$ (right) as a function of $\kappa$. Blue markers represent dCS solutions, while orange markers show the sGB solutions. At the top or bottom of either panel, the spectral order of the solutions is listed. }
    \label{fig:ErrorResidual}
\end{figure*}

We validate our numerical results by verifying the numerical convergence of our solution. 
To do so, we define the backward modulus differences as the $L^{\infty}$-norm of the difference in $\vartheta$ or $H_i$ between successive spectral orders
\begin{equation}
\begin{split}
    \mathcal{B}_{\vartheta}(N_z, N_{\chi}) &= \max_{r, \chi} \Big|\vartheta^{[N_z, N_{\chi}]} - \vartheta^{[N_z, N_{\chi}-2]}\Big| \\
    \mathcal{B}_T({\cal N}_z, {\cal N}_{\chi}) &= \sum_{i = 1}^{4} \max_{r, \chi} \Big|H_i^{[{\cal N}_z, {\cal N}_{\chi}]} - H_i^{[{\cal N}_z - 10, {\cal N}_{\chi}]}\Big|
\end{split}
\end{equation}
as the measure of convergence, where the superscripts $N_{z, \chi}$ and ${\cal N}_{z, \chi}$ indicate the spectral order of the solution \cite{Lam:2025elw, Lam:2025fzi, Chung:2026zon}. 
In the left panel of Fig.~\ref{fig:BMD}, we plot ${\cal B}_{\vartheta}$ for the set of solutions with the largest spin that we have computed. 
Both scalar field solutions converge exponentially, as expected from the pseudospectral method. 
Similarly, ${\cal B}_T$ also decays approximately exponentially. 
At the largest ${\cal N}_z$, the backward modulus difference is ${\cal B}_T \approx 10^{-3}$. 
We find that this precision is enough to study the near-extremal behavior of BHs in these theories. 
We omit the convergence plots for solutions with smaller spins because they show the same behavior, and they have a lower ${\cal B}_{\vartheta}$ and ${\cal B}_T$. 
These results confirm that the solutions have converged. 

We also investigate the accuracy of our solutions. 
To do so, we define the error residual of the scalar field equation as 
\begin{equation}
    {\cal E}_{\vartheta} = \max_{r, \chi} |E_{\vartheta}|.
\end{equation}
We plot the error residual as a function of $\kappa$ in the left panel of Fig.~\ref{fig:ErrorResidual}. 
Each shaded region represents ${\cal E}_{\vartheta}$ computed with $\vartheta$ at the spectral order listed at the top or bottom of the panel. 
In each shaded region, ${\cal E}_{\vartheta}$ increases as we decrease $\kappa$. 
For all solutions considered, we keep ${\cal E}_{\vartheta} < 10^{-7}$ to ensure that the scalar field is accurate and does not affect the accuracy of the metric corrections.

Next, to gauge the accuracy of the $H_i$ corrections, we define an error measure by the $L_2$ norm of the residual of Eq.~\eqref{eq:SpectralMatrixEq}, 
\begin{equation}\label{eq:L2Error}
    {\cal E} = \big|\big|\tilde{\mathbb{D}}{\bf v} - \tilde{\bf s}\big|\big|_2.
\end{equation}
The residual $\cal E$ can be interpreted as a weighted combination of the field equations, which would be exactly zero if the solutions were exact. 
In the right panel of Fig.~\ref{fig:ErrorResidual}, we show $\cal E$ as a function of $\kappa$. 
For each $\kappa$, we compute $\cal E$ using solutions with the maximum ${\cal N}_z$ and ${\cal N}_{\chi}$ that is available. 
In dCS gravity, similar to the case for ${\cal E}_{\vartheta}$, the residual $\cal E$ generally increases in each shaded region. 
When $\kappa \approx 10^{-4}$, the residual ${\cal E} \lesssim 10^{-7}$. 
For sGB gravity, the trend is similar, although the residual becomes slightly larger. 

\section{\MakeLowercase{d}CS Black Holes} \label{sec:dCS}

In this section, we present the dCS BH solutions constructed with the horizon-adapted formalism.
We first study the near-extremal scalar field, metric corrections, and conserved charges, and then construct the extremal solution directly.
We finally take the near-horizon limit of this extremal exterior and compare it with the dCS-deformed NHEK geometry obtained from the enhanced-symmetry approach.

\subsection{Sub-extremal Solutions}

\begin{figure}[t!]
    \centering
    \includegraphics[width=\columnwidth]{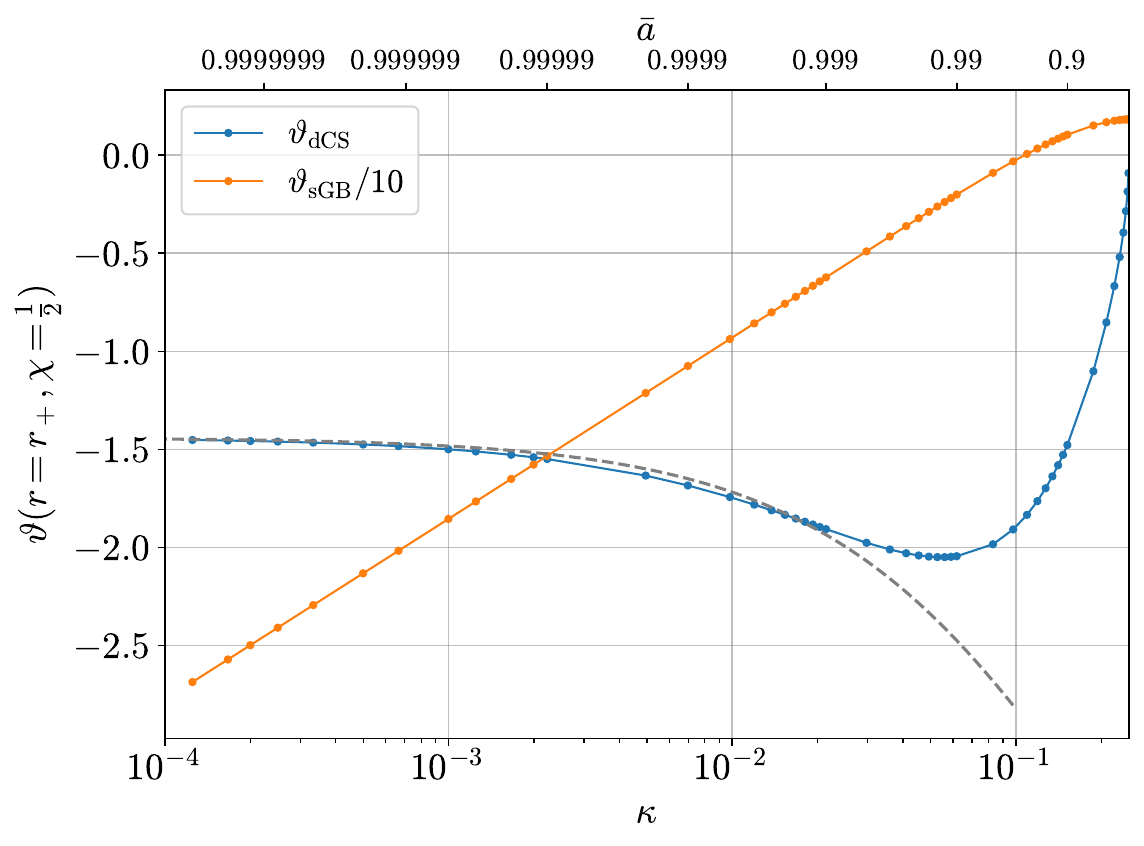}
    \caption{dCS and sGB scalar fields $\vartheta(r = r_+, \chi = \frac{1}{2})$ against $\kappa$. The sGB scalar field is scaled by a factor of $1/10$ to match the scale of the dCS solution. The gray curve is the prediction of $\vartheta_{\rm dCS}(r_+,1/2)$ derived from the analytic extremal solution \cite{McNees:2015srl}, which shows that the $\kappa\log\kappa$ dependence comes from the $(r-1)\log(r-1)$ radial dependence in the extremal scalar field. As $\kappa \to 0$, the gray curve approaches $\vartheta_{\rm dCS}$.}
    \label{fig:Scalar_Field_Horizon}
\end{figure}

Let us first study the profile of the dCS pseudoscalar field. 
As $\kappa$ decreases, the scalar field develops higher multipole moments, as found in \cite{Lam:2025fzi}. 
At spatial infinity, these Legendre modes decay as $r^{-l-1}$, as required by the spectral ansatz in Eq.~\eqref{eq:Scalar_Field_Ansatz}. 
At the horizon, we plot the scalar field magnitude in Fig.~\ref{fig:Scalar_Field_Horizon} as a function of $\kappa$. (For comparison, the plots in this section also contain the results for sGB,  but we defer discussion of them until Sec. \ref{sec:sGB}.)
As we lower $\kappa$, the value of the scalar field at the horizon converges to a finite value, with magnitude matching the analytical calculations of \cite{McNees:2015srl}. 
Since the extremal limit appears continuous, we expect the analytic extremal solution can be used as a good approximation to the near-extremal solutions. 
From \cite{McNees:2015srl}, the extremal solution has $\vartheta_{\rm dCS} - \vartheta_{\rm dCS, ext} \sim (r - 1) \log (r - 1)$, where $\vartheta_{\rm dCS, ext}$ is the value of the scalar field at extremality. 
Evaluating at the near-extremal horizon, we have $\vartheta_{\rm dCS} - \vartheta_{\rm dCS, ext} \sim (r_+ - 1) \log (r_+ - 1) \sim \kappa \log \kappa$. 
The $\kappa \log \kappa$ dependence matches the convergence behavior near $\kappa = 0$ shown in Fig.~\ref{fig:Scalar_Field_Horizon}. 

\begin{figure*}[t!]
    \centering
    \includegraphics[width=\textwidth]{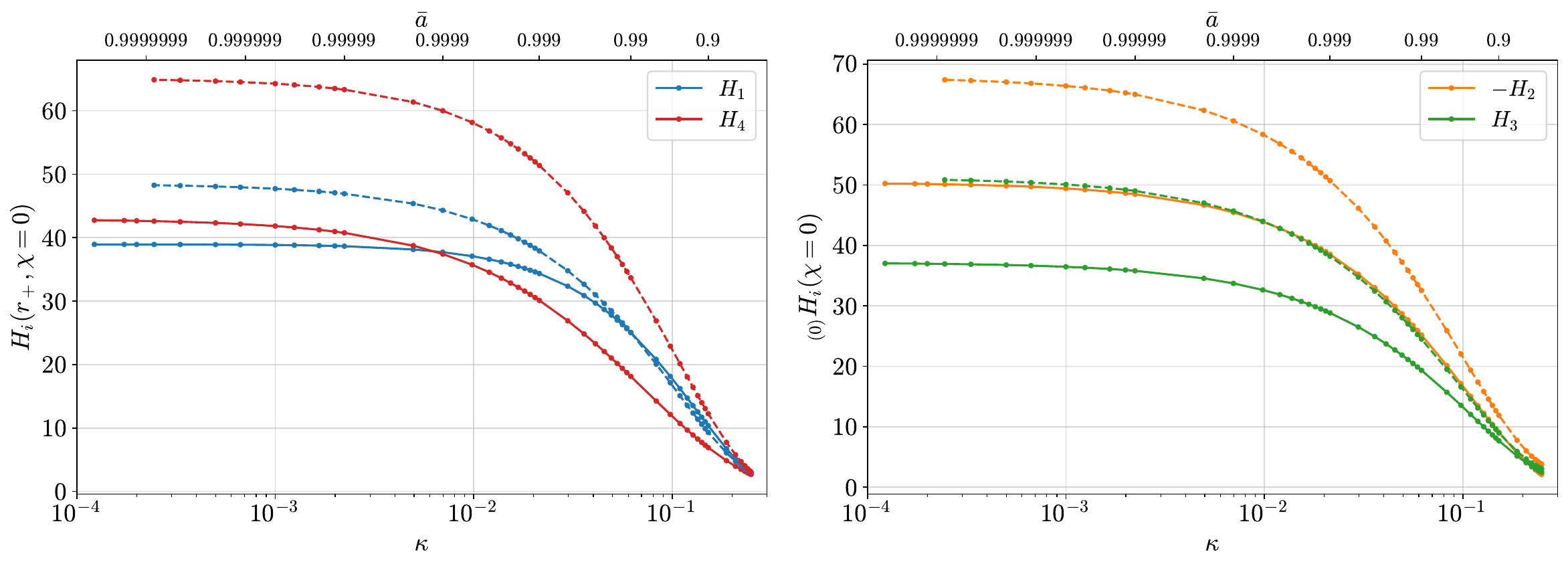}
    \caption{Metric corrections of dCS and sGB BHs $H_i(r,\chi=0)$ against $\kappa$ at the horizon (left) and at spatial infinity (right). dCS corrections are depicted by solid lines, while sGB corrections are given in dashed lines. At the horizon $H_1 = H_3$ and $H_2 = 0$, thus we refrain from plotting $H_{2,3}$; similarly at spatial infinity, $H_1 = 0$ and $H_3 = H_4$, hence we do not plot $H_{1,4}$. Note that ${}_{(0)}H_i(\chi) = H_i(r \to \infty, \chi)$ by the definition in Eq.~\eqref{eq:H_Inf_Expansion}. }
    \label{fig:Hi_kappa}
\end{figure*}

\begin{figure}[t!]
    \centering
    \includegraphics[width=\columnwidth]{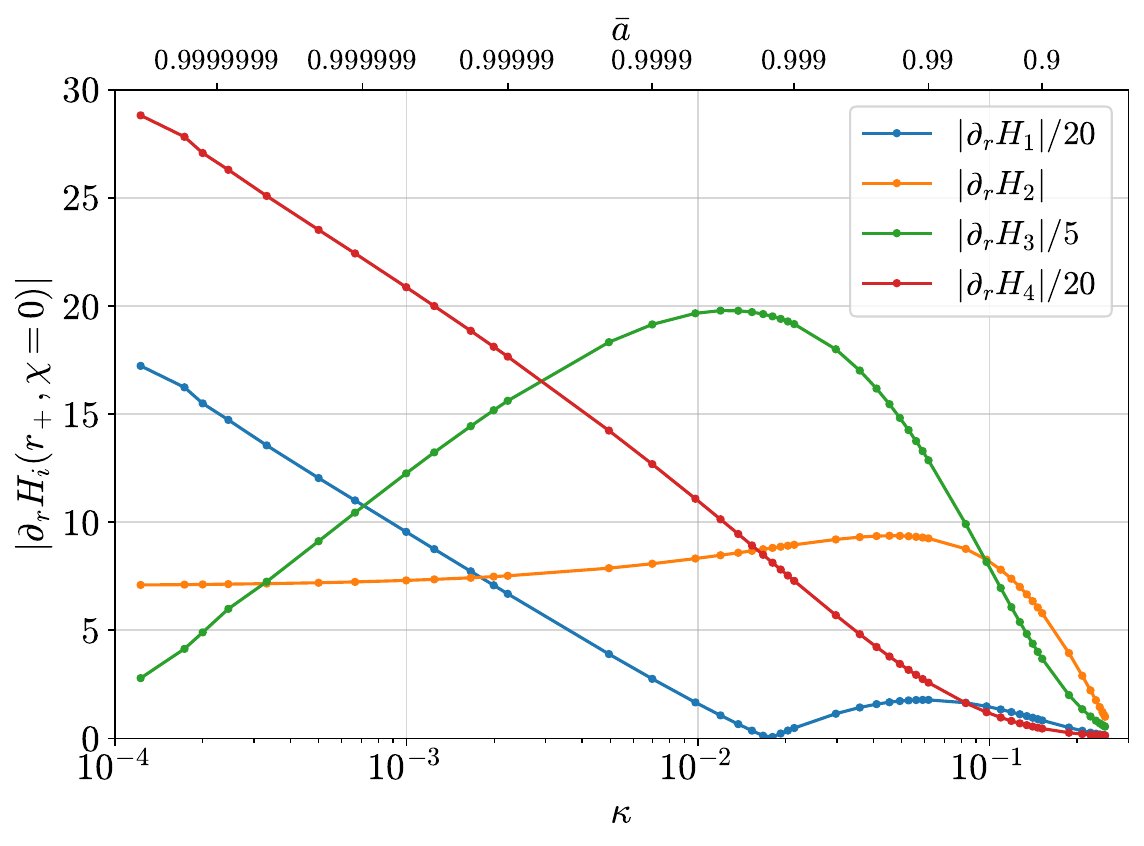}
    \caption{Derivative of the dCS metric corrections $|\partial_r H_i(r = r_+, \chi = 0)|$. Note that individual lines are rescaled such that they appear on the plot with similar magnitudes. }
    \label{fig:DrHi_kappa_dCS}
\end{figure}

Next, we investigate the behavior of the metric corrections at the horizon and at spatial infinity. 
Using solutions at different $\kappa$, we plot $H_{1}(r_+, 0)$ and $H_{4}(r_+, 0)$ against $\kappa$ in the left panel of Fig.~\ref{fig:Hi_kappa}.
We do not plot $H_{2}(r_+,0)$ or $H_{3}(r_+,0)$, since $H_3(r_+, 0)$ is identical to $H_1(r_+,0)$ and $H_2(r_+, 0) =0$, a consequence of Eq.~\eqref{eq:HorizonBC}. 
For dCS corrections, as $\kappa$ becomes smaller, both $H_{1}$ and $H_{4}$ increase rapidly at first, then flatten as $\kappa$ reaches around $10^{-3}$. 
From this trend, we expect that $H_i$ will converge to the extremal solution as $\kappa \to 0$, as we will show later in this section. 
At spatial infinity, the right panel of Fig.~\ref{fig:Hi_kappa} shows that $H_{2}$ and $H_{3}$ also tend to finite values that will be related to the ADM mass and the ADM angular momentum.

Let us next examine the radial derivatives of the metric corrections at the horizon, $\partial_r H_i(r_+,0)$.
As shown in Fig.~\ref{fig:DrHi_kappa_dCS}, $|\partial_r H_2|$ approaches a finite value, approximately $7.1$, while the derivatives of the remaining metric functions grow logarithmically as $\kappa \to 0$.
This means that the extremal dCS solution is not completely smooth in these coordinates. The metric is continuous, but not differentiable, at the horizon. As we will show in Sec.~\ref{sec:BreakdownEFT}, this logarithmic behavior produces divergent tidal forces, but does not produce scalar curvature singularities.

The origin of the logarithm can be traced to the near-horizon behavior of the dCS pseudoscalar field.
As discussed above, near extremality $\vartheta_{\rm dCS}$ contains terms proportional to $(r-1)\log(r-1)$.
These terms are finite at the horizon, but their radial derivatives diverge logarithmically.
When inserted into the source tensor, for example through $[\mathscr{S}_{r}{}^{r}]^{(0)}\sim \Delta(\partial_r\vartheta_{\rm dCS})^2$, this structure propagates to the metric corrections and produces terms of the same schematic form, $H_i\sim (r-1)\log(r-1)$.
Evaluated on the near-extremal horizon, their radial derivatives scale as $\log(r_+-1)\sim\log\kappa$.

Finally, we examine the conserved charges, namely, the ADM mass and the ADM angular momentum. 
In our metric ansatz, the radial coordinate $r$ does \textit{not} coincide with the areal radius $r_{\rm areal}$, which geometrically is the radius that corresponds to 2-spheres at spatial infinity \cite{Cano_Ruiperez_2019}. 
The two radii are related by the following equation:
\begin{equation}\label{eq:Areal_Radius}
    r = r_{\rm areal} \left(1 - \zeta \frac{{}_{(0)}H_3}{2}\right) - \zeta \frac{{}_{(1)}H_3}{2} + O\left(\frac{1}{r}\right).
\end{equation}
Writing the metric in terms of $r_{\rm areal}$ and expanding at large $r_{\rm areal}$, we can extract the ADM mass and angular momentum directly from the asymptotic metric, 
\begin{equation} \label{eq:ADM}
\begin{split}
    M &= 1 + \frac{\zeta}{2}\Big[{}_{(0)}H_3 + {}_{(1)}H_3\Big], \\
    J &= \bar{a} \left[1 + \frac{\zeta}{2}\left(2 {}_{(0)}H_2 + {}_{(0)}H_3 + 2 {}_{(0)}H_4\right)\right].
\end{split}
\end{equation}
These formulae can equivalently be derived using Komar integrals, since the spacetime is stationary and axisymmetric. 

\begin{figure}[t!]
    \centering
    \includegraphics[width=\columnwidth]{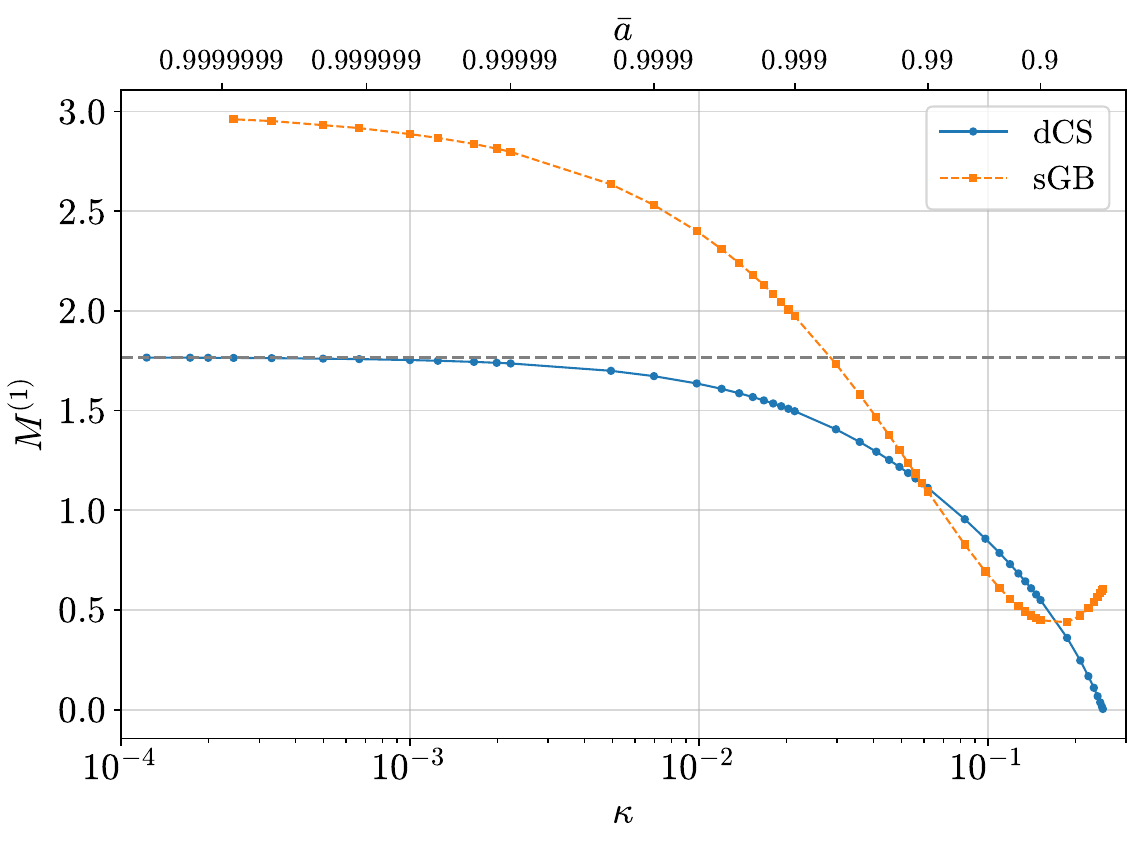}
    \caption{EFT corrections to the ADM mass $M^{(1)}$ in dCS gravity (blue, solid) and in sGB gravity (orange, dashed) BHs.
    As $\kappa \to 0$, both $M^{(1)}$ tend to finite values. 
    Moreover, $M_{\rm dCS}^{(1)}$ matches the mass correction in the extremal limit, represented by the horizontal dashed line. }
    \label{fig:deltaM}
\end{figure}

\begin{figure}[t!]
    \centering
    \includegraphics[width=\columnwidth]{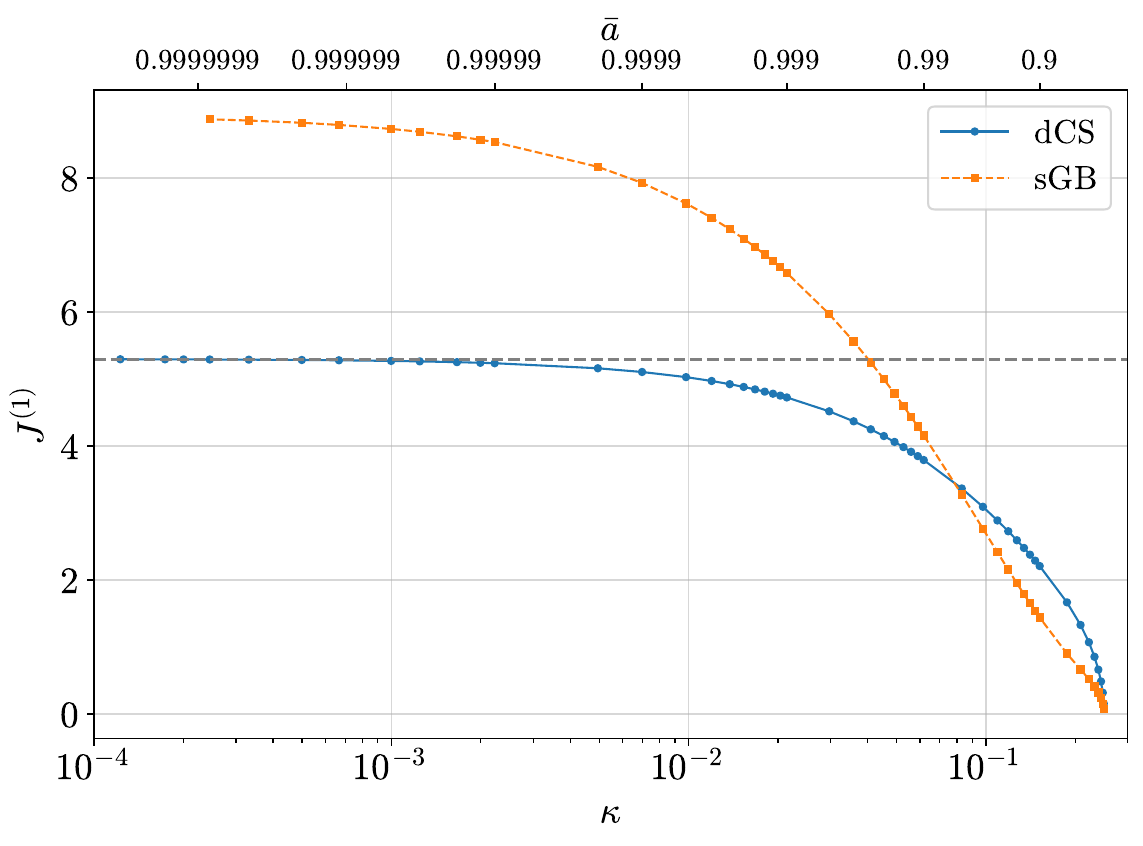}
    \caption{Identical to Fig.~\ref{fig:deltaM}, except we plot the EFT corrections to the ADM angular momentum $J^{(1)}$. 
    }
    \label{fig:deltaJ}
\end{figure}

We plot the EFT-correction to the ADM mass $M^{(1)}$ in Fig.~\ref{fig:deltaM}, and the EFT-correction to the ADM angular momentum $J^{(1)}$ in Fig.~\ref{fig:deltaJ}. 
Observe that both $M^{(1)}$ and $J^{(1)}$ in dCS gravity are positive over the full range of $\kappa \in [0, 1/4]$, leading to an increase in total mass and angular momentum ($\zeta$ is always positive). 
In contrast, in other higher-derivative gravity theories \cite{Reall:2019sah}, $M^{(1)}$ can be either positive or negative for different values of $\kappa$.
When $\kappa = 1/4$ ($\bar{a} = 0$), $M^{(1)} = 0 = J^{(1)}$. 
This is because the Schwarzschild solution is also a solution in dCS gravity, so neither the ADM mass nor the ADM angular momentum are shifted. 
As $\kappa$ decreases, $M^{(1)}$ steadily increases and approaches a constant value of around $1.8$, while $J^{(1)}$ increases monotonically to approximately $5.3$. 
All of these results suggest the extremal limit is continuously connected to the subextremal family, and can be attained using the same methods, which we now present below. 

\subsection{Extremal Solution}

Let us now construct the extremal dCS solution with pseudospectral methods.
The metric equations require the scalar field as an input.
For the extremal solution, however, we do not compute this scalar field with the pseudospectral expansion of Sec.~\ref{subsec:Scalar_Field}.
Instead, we use the analytic series solution of~\cite{McNees:2015srl}.
The reason is that the extremal dCS pseudoscalar contains terms proportional to $(r-1)\log(r-1)$, which means that while the scalar itself is finite at the horizon, its radial derivative is not.
A finite pseudospectral expansion can approximate this behavior, but the analytic series builds the logarithmic terms in explicitly and is, therefore, more efficient and more accurate at fixed truncation order.
As a check, the pseudospectral and analytic scalar-field solutions agree within numerical error.

The numerical procedure to compute the extremal metric corrections is almost the same as that in the sub-extremal case. 
Below, we list the modifications and numerical implementations.
\begin{enumerate}
    \item To compute the source tensor $\mathscr{S}_{\mu}{}^{\nu}$, as stated before, we use the analytic extremal scalar field solution up to $P_{l = 47}(\chi)$ \cite{McNees:2015srl}. 
    We find that increasing the number of terms included would not significantly affect the accuracy of the final solution.
    \item We use the same set of boundary conditions as before, with two additional ones that improve the accuracy of the solution: 
    \begin{equation}\label{eq:Extremal_BC}
    \begin{split}
        \left[H_2 - \frac{\partial H_2}{\partial r}\right]_{r = r_+} &= \frac{1972 + 315 \pi}{420} \\
        \partial_{\chi} H_2(r_+, \chi) &= 0
    \end{split}
    \end{equation}
    The first boundary condition fixes the coefficient of the NHEK fibration to be the dCS-deformed value $P_2$, which will be presented in the next subsection. Together with the horizon condition $H_2(r_+,\chi)=0$, it implies $-\partial_r H_2(r_+,\chi)=P_2=(1972+315\pi)/420$. Thus, the magnitude $|\partial_r H_2(r_+,\chi)|$ approaches the value shown in Fig.~\ref{fig:DrHi_kappa_dCS}. The second condition fixes $\Omega_H$ to be $\chi$-independent, as required by the rigidity theorem \cite{Poisson:2009pwt}. 
    \item We use ${\cal N}_z = 180$ and ${\cal N}_{\chi} = 32$. 
    The extremal solution converges exponentially with ${\cal B}_T \approx 10^{-3}$, in approximately the same way as the near-extremal solution. 
    The error residual is ${\cal E} \approx 4 \times 10^{-9}$. 
\end{enumerate}

With the pseudospectral solution, we compute the EFT corrections to the ADM mass and angular momentum of the extremal solution and find them to be
\begin{equation}
    M^{(1)} = 1.77, \quad J^{(1)} = 5.30.
\end{equation}
We plot these two values in Fig.~\ref{fig:deltaM} and \ref{fig:deltaJ} as horizontal grey dashed lines, and these corrections match the limiting values computed using the sub-extremal solutions. 
This shows that the extremal solutions are connected to the sub-extremal ones. 

Let us comment on past attempts to construct extremal BH spacetimes in dCS gravity. 
Earlier studies of (near-) extremal dCS metric corrections focused primarily on the trace $g^{\mu\nu}_{(0)}g_{\mu\nu}^{(1)}$ in the Lorenz gauge \cite{McNees:2015srl, Stein:2014xba}. 
This is because the equations for the trace reduce to a Poisson-type equation in the Kerr background and can be solved either analytically or numerically using the same method as the scalar field. 
In this gauge, the trace of the extremal solution diverges logarithmically at the unperturbed horizon \cite{McNees:2015srl}. 
However, the Lorenz gauge is not horizon-locking, and thus, based on slowly-rotating results, the authors of \cite{McNees:2015srl} argued that the horizon would shift to a slightly larger radius, and the singularity would be hidden behind the perturbed horizon. 
It was therefore conjectured that the horizon singularity is not a physical singularity, and our results confirm this conjecture. 
We find that the trace of the metric perturbation remains finite at the horizon, as do all metric components. 
The previously observed divergence arises entirely from the choice of gauge.
In particular, our results suggest that the Lorenz gauge is incompatible with the extremal limit, in the sense that it fails to provide a regular description of the horizon geometry in this regime.

\subsection{Near-horizon Extremal Geometry}

In GR, since an extremal horizon is infinitely far away along a stationary surface, one can define a limiting near-horizon geometry which has enhanced symmetry. 
Since the full dCS extremal solution can be numerically constructed, one can ask if it has a similar near-horizon geometry. 
Before diving into the dCS case, let us briefly review the near-horizon extremal Kerr (NHEK) metric and some related properties.

To obtain the NHEK metric, one zooms into the near-horizon region through the following transformation: 
\begin{align}\label{eq:NHEKTransformation}
    t = 2\tilde{t} / \epsilon, \quad 
    r = 1 + \epsilon \, \tilde{r}, \quad 
    \phi = \tilde{\phi} + 2 \Omega_H \tilde{t}/\epsilon, 
\end{align}
with $\Omega_H = 1/2$ and $\epsilon \to 0$ \cite{Bardeen:1999px}. 
The resulting NHEK metric becomes
\begin{align}
    ds_{\rm NHEK}^2 &= \lim_{\epsilon \to 0} \lim_{a \to 1} g_{\mu \nu}^{(0)} dx^{\mu} dx^{\nu} \nonumber \\
    &= (1 + \chi^2) \left(-\tilde{r}^2 d\tilde{t}^2 + \frac{d\tilde{r}^2}{\tilde{r}^2} + \frac{d\chi^2}{1 - \chi^2}\right) \nonumber \\
    &+ \frac{4(1 - \chi^2)}{1 + \chi^2} (d\tilde{\phi} + \tilde{r} d\tilde{t})^2. 
\end{align}
The isometry group in Kerr is $\mathbb{R} \times U(1)$, which is generated by the Killing vectors $\partial_t$ and $\partial_{\phi}$. In the near-horizon regime, this isometry is now promoted to $SO(2,1) \times U(1)$, with the following Killing vectors~\cite{Bardeen:1999px}. 
\begin{equation}\label{eq:AdS2KillingVectors}
\begin{split}
    \xi_0 &= \tilde{t} \, \partial_{\tilde{t}} - \tilde{r} \, \partial_{\tilde{r}}, \\
    \xi_+ &= \partial_{\tilde{t}}, \\
    \xi_- &= \left(\tilde{t}^2 + \frac{1}{\tilde{r}^2}\right) \partial_{\tilde{t}} - 2 \, \tilde{t} \, \tilde{r}  \, \partial_{\tilde{r}} - \frac{2}{\tilde{r}} \partial_{\tilde{\phi}}, \\
    \xi_{\tilde{\phi}} &=\partial_{\tilde{\phi}}\,.
\end{split}
\end{equation}
The Killing vector $\xi_+$ represents $\tilde{t}$-translation symmetry, while the Killing vector $\xi_0$ can be interpreted as a dilation symmetry. The latter implies that the NHEK metric is invariant under the scalings $\tilde{t} \mapsto \tilde{t}/\alpha$ and $\tilde{r} \mapsto \alpha \tilde{r}$ for any $\alpha \in \mathbb{R}$.

Extending this analysis to dCS gravity, the near-horizon solution can be found by considering isometry-preserving deformations of the NHEK metric \cite{Chen:2018jed, Cano:2023dyg}.
First, in the NHEK background, the scalar field equation in Eq.~\eqref{eq:SF0} can be solved exactly to find\footnote{Due to a different convention of the coupling constant, our scalar fields are -8 times those computed in \cite{Chen:2018jed}. }
\begin{equation}\label{eq:NHEK_SF_dCS}
\begin{split}
    \vartheta_{\text{dCS-NHEK}}(\chi) &= \frac{2\chi(\chi^4 + 2\chi^2 - 7)}{(1 + \chi^2)^3} + 4\arctan\chi, 
\end{split}
\end{equation}

With the scalar field solution in hand, since the source tensor $\mathscr{S}_{\mu}{}^{\nu}$ also has the enhanced symmetries of the $SO(2, 1) \times U(1)$ group, the corrected metric also has this same isometry group. 
Using this fact, the near-horizon solution can be found by solving Eq.~\eqref{eq:EE1} in the NHEK background  \cite{Chen:2018jed, Cano:2023dyg}. 
With the following metric ansatz, 
\begin{widetext}
\begin{equation}
\begin{split}
    ds^2 &= [1 + \zeta P_1(\chi)] (1 + \chi^2) \left(-\tilde{r}^2 d\tilde{t}^2 + \frac{d\tilde{r}^2}{\tilde{r}^2}\right) + [1 + \zeta P_3(\chi)] \frac{1 + \chi^2}{1 - \chi^2} d\chi^2 + 4 [1 + \zeta P_4(\chi)] \frac{1 - \chi^2}{1 + \chi^2} \left[d\tilde{\phi} + (1 + \zeta P_2) \tilde{r} d\tilde{t}\right]^2, 
\end{split}
\end{equation}
the solution is found analytically to be\footnote{The dCS-deformed NHEK solution given in \cite{Chen:2018jed} contains a conical singularity. This is because they forced one of the Killing vectors to be the same as that in Kerr, which set $P_2 = 0$ [see Eq.~\eqref{eq:xi_minus}].  As in \cite{Cano:2023dyg, Cano:2024bhh}, we instead demand that the solution be free of conical singularities. This fixes $P_2$ to be the value we state in Eq.~\eqref{eq:dCS_NHEK_sol}. }
\begin{equation}\label{eq:dCS_NHEK_sol}
\begin{split}
    P_1(\chi) &= c_1 + g_1 G(\chi) - \frac{6 \chi \arctan(\chi)}{1 + \chi^2} + \frac{62917 + 201207 \chi^2 + 1201250 \chi^4 + 1181774 \chi^6 + 622457 \chi^8 + 128859 \chi^{10}}{1680 (1 + \chi^2)^6}, \\
    P_2 &= \frac{1972 + 315 \pi}{420}, \\
    P_3(\chi) &= P_1(\chi) - \frac{975}{512\sqrt{2}} \left(g_1 + \frac{\sqrt{2}(32 - g_1)}{1 + \chi^2}\right), \\
    P_4(\chi) &= c_1 - 2P_2 - \frac{2 g_1}{1 - \chi^2} G(\chi) + \frac{6\chi \arctan(\chi)}{1 + \chi^2} \\
    &+ \frac{20917 + 404172 \chi^2 - 195845 \chi^4 + 225840 \chi^6 + 371659 \chi^8 + 246324 \chi^{10} + 59061 \chi^{12}}{1680 (1 + \chi^2)^6}, 
\end{split}
\end{equation}
\end{widetext}
where $c_1$ and $g_1$ are integration constants, and 
\begin{align}
    G(\chi) &= -\frac{975 \chi \sqrt{1 - \chi^2}}{512 \sqrt{2} (1 + \chi^2)} \nonumber \\
    &\times \left[\arcsin(\chi) - \arctan\left(\frac{\sqrt{2}\,\chi}{\sqrt{1 - \chi^2}}\right)\right].
\end{align}
These functions $P_i(\chi)$ should not be confused with the Legendre polynomials. 
Using the same notation as in \cite{Chen:2018jed}, we will refer to this solution as the dCS-deformed NHEK solution. 

The local dCS-deformed NHEK solution contains two remaining constants.
The constant $c_1$ is not fixed by the local near-horizon equations; it corresponds to the overall normalization of the throat and must be fixed by matching to the full asymptotically flat extremal exterior.
We perform this matching below.
The constant $g_1$, on the other hand, reflects a residual gauge freedom associated with an infinitesimal coordinate transformation in $\chi$ \cite{Cano:2024bhh}.
We fix this gauge by choosing $g_1 = 32$, which puts the solution in the attractor form \cite{Astefanesei:2006dd}.
With this choice, $P_1(\chi)$ and $P_3(\chi)$ differ only by the constant $-975/(16\sqrt{2})$, which is precisely $\Lambda_{\rm dCS}$.
This suggests that the dCS-deformed NHEK solution and the full exterior ansatz are compatible, since the same constant is already built into Eq.~\eqref{eq:metric} through $\Xi(r)$.

The deformation proportional to $P_2$ also changes the explicit form of one of the $SO(2,1)$ generators.
For the Kerr NHEK metric, the special conformal Killing vector was already presented in Eq.~\eqref{eq:AdS2KillingVectors}. In the dCS-deformed NHEK geometry, the same generator is instead represented by
\begin{equation}\label{eq:xi_minus}
    \xi_- =
    \left(\tilde{t}^2 + \frac{1}{\tilde{r}^2}\right)\partial_{\tilde{t}}
    -2\tilde{t}\tilde{r}\partial_{\tilde{r}}
    -\frac{2(1+\zeta P_2)}{\tilde{r}}\partial_{\tilde{\phi}} .
\end{equation}
The change in the $\partial_{\tilde{\phi}}$ component compensates for the corrected fibration
$d\tilde{\phi}+(1+\zeta P_2)\tilde{r}d\tilde{t}$ in the metric ansatz.
Thus, the deformation changes the representation of the Killing vector, but not the isometry group itself, because the geometry still has the same $SO(2,1)\times U(1)$ symmetry as NHEK.

\begin{figure*}[tbh]
    \centering
    \includegraphics[width=\linewidth]{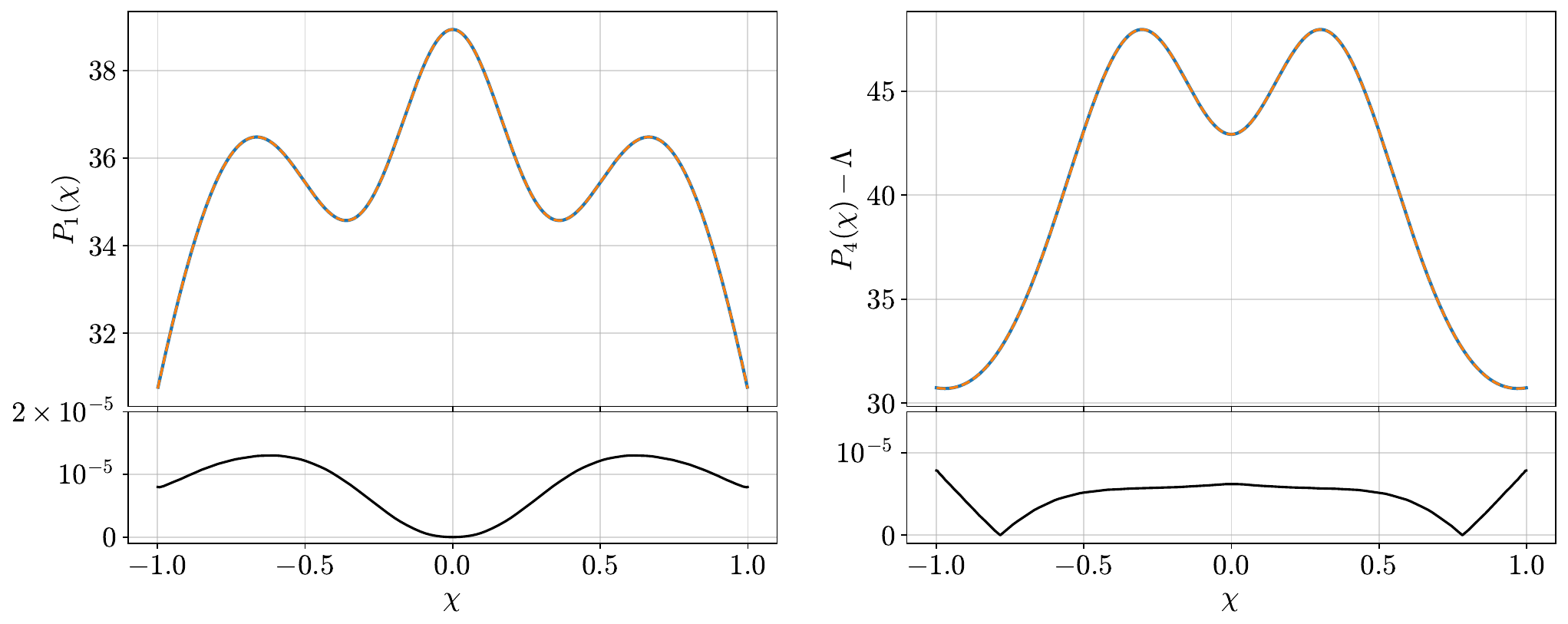}
    \caption{Exterior metric corrections evaluated at the horizon $H_i(r_+, \chi)$ and dCS-deformed NHEK solution $P_i(\chi)$. Left: Top panel shows $P_1(\chi)$ (blue) and $H_1(\chi)$ (orange), with the relative error $|P_1 - H_1| / |P_1|$ shown in the bottom panel. Right: Top panel shows $P_4(\chi) - \Lambda$ (blue) and $H_4(\chi)$ (orange), with relative error $|(P_4 - \Lambda) - H_4| / |P_4 - \Lambda|$ shown in the bottom panel. }
    \label{fig:NHEK_Match}
\end{figure*}

We can now compare this analytic dCS-deformed NHEK solution with the near-horizon limit of the full asymptotically flat extremal dCS exterior constructed in the previous subsection.
To do so, we zoom into the horizon of the numerical exterior solution and use the corrected horizon angular velocity in the near-horizon coordinate transformation.
The appropriate near-horizon limit is therefore obtained through the transformation
\begin{equation}\label{eq:modifiedNHEKTransformation}
    t = 2\tilde{t} / \epsilon, \quad 
    r = 1 + \epsilon \tilde{r}, \quad 
    \phi = \tilde{\phi} + 2 \Omega_H \tilde{t}/\epsilon, 
\end{equation}
as $\epsilon \to 0$, where $\Omega_H = \Omega_H^{(0)} \left[1 + \zeta H_2(r_+, \chi)\right]$. Recalling that $H_2(r_+, \chi)$ is a constant in $\chi$, as imposed by the boundary condition and the rigidity theorem, and taking the $\epsilon \to 0$ limit, Eq.~\eqref{eq:metric} transforms to
\begin{equation}
\begin{split}
    ds^2 &= -[1 + \zeta H_1(r_+, \chi)] (1 + \chi^2) \tilde{r}^2 d\tilde{t}^2 \\
    & \quad + [1 + \zeta H_3(r_+, \chi)] (1 + \chi^2) \frac{d\tilde{r}^2}{\tilde{r}^2} \\
    & \quad + [1 + \zeta (H_3(r_+, \chi) + \Lambda)] \frac{1 + \chi^2}{1 - \chi^2} d\chi^2 \\
    & \quad + 4 [1 + \zeta (H_4(r_+, \chi) + \Lambda)] \frac{1 - \chi^2}{1 + \chi^2} \\
    & \quad \times \bigg\{d\tilde{\phi} + \Big[1 + \zeta \Big(H_2(r_+, \chi) - \partial_r H_2(r_+, \chi)\Big)\Big] \tilde{r} d\tilde{t}\bigg\}^2.
\end{split}
\end{equation}
If the near-horizon limit is to match the dCS-deformed NHEK metric, the following must be true:
\begin{equation} \label{eq:P-conditions}
\begin{split}
    P_1(\chi) &= H_1(r_+, \chi) = H_3(r_+, \chi), \\
    P_2 &= H_2(r_+, \chi) - \partial_r H_2(r_+, \chi), \\
    P_4(\chi) &= H_4(r_+, \chi) + \Lambda. 
\end{split}
\end{equation}

Let us now separate the relations in Eq.~\eqref{eq:P-conditions} into those imposed by construction and those that provide genuine checks of the matching.
The equality $H_1(r_+,\chi)=H_3(r_+,\chi)$ in the first line follows from the horizon boundary condition in Eq.~\eqref{eq:HorizonBC}.
Similarly, the second line follows from the extremal boundary condition in Eq.~\eqref{eq:Extremal_BC}, together with $H_2(r_+,\chi)=0$.
The remaining matching is therefore in the first and third lines. 
We must check whether $P_1(\chi)$ agrees with $H_1(r_+,\chi)$ and whether $P_4(\chi)$ agrees with $H_4(r_+,\chi)+\Lambda$ for all $\chi$, after we fix the remaining integration constant $c_1$.

We fix $c_1$ by matching one value, which we choose to be the equatorial value,
\begin{equation}
    P_1(0)=H_1(r_+,0).
\end{equation}
Since $P_1(0)=c_1+62917/1680$, the matching condition gives
\begin{equation}
    c_1 = H_1(r_+,0)-\frac{62917}{1680} \approx 1.49 ,
\end{equation}
where the final value is obtained from the numerical extremal exterior solution.
Figure~\ref{fig:NHEK_Match} shows that the resulting $P_1(\chi)$ and $P_4(\chi)-\Lambda$ agree with the corresponding horizon values of the full exterior solution at the level of $10^{-5}$.
This numerical agreement verifies that the full asymptotically flat extremal dCS solution approaches the isometry-preserving dCS-deformed NHEK throat.
Equivalently, in this theory, the local dCS-deformed NHEK solution admits an asymptotically flat exterior extension, which is not guaranteed for BHs in higher-derivative gravity \cite{Cano:2019ozf}.

\section{\MakeLowercase{s}GB Black Holes} \label{sec:sGB}

In this section, we study the corresponding sGB BH solutions and their approach to extremality.
The finite-$\kappa$ solutions are regular, but the $\kappa \to 0$ limit behaves differently from dCS gravity.
We first trace this difference to the scalar field, then examine the metric corrections and conserved charges, and finally show why the sGB-deformed NHEK solution does not describe the near-horizon region of the asymptotically flat solution.

\subsection{Scalar Field}

Let us start with the scalar field. 
In Fig.~\ref{fig:Scalar_Field_Horizon}, the sGB scalar field at the horizon decreases logarithmically with $\kappa$. 
Unlike in dCS gravity, we then expect that $\vartheta_{\rm sGB}$ will be unbounded at the horizon in the extremal limit. 
To explain the $\log \kappa$ behavior of the near-extremal sGB scalar field, let us first discuss the extremal scalar field solution. 
To solve the extremal scalar field equation $E_{\vartheta}$ [Eq.~\eqref{eq:SF0}], we first note that it admits an eigenfunction expansion in Legendre modes \cite{McNees:2015srl, Berti:2018cxi}.
For each Legendre mode, we have 
\begin{equation}\label{eq:SFE_Legendre}
    \frac{d}{dr}\left[(r-1)^2 \frac{d\vartheta_l}{dr}\right]-l(l+1)\vartheta_l=s_l(r), 
\end{equation}
where $s_l(r)$ can be found in Eq.~\eqref{eq:s_ell}.
We are particularly interested in the monopole term $(l = 0)$ here. The Legendre-mode-decomposed equation near the horizon reads
\begin{equation}
\label{eq:d-drho}
    \frac{d}{d\rho}\left[\rho^2\frac{d\vartheta_{l = 0}}{d\rho}\right]=4+O(\rho),
\end{equation}
where $\rho \equiv r - 1$, so that $\rho = 0$ is the extremal Kerr horizon.
Integrating twice, we find 
\begin{equation}\label{eq:sGB_SF_l0}
    \vartheta_{l = 0}(\rho) = 4\log\rho + O(1). 
\end{equation}
When evaluated at the extremal horizon $\rho = 0$, the scalar field diverges.  
This agrees with the limiting behavior shown in Fig.~\ref{fig:Scalar_Field_Horizon}. 
The full exterior solution can, in fact, be found in the same way, without a near-horizon expansion. 
In Appendix~\ref{sec:sGB_Scalar_Field}, we solve Eq.~\eqref{eq:SFE_Legendre} analytically, and provide the first two terms in the Legendre mode expansion. 
Moving to near-extremal solutions, we expect this logarithmic term to be the leading-order contribution.  
Then, at the horizon, $\vartheta_{\rm sGB} \sim \log (r_+ - 1) \sim \log \kappa$, which matches what we observe numerically. 

This divergence of the extremal scalar field is not new, and in fact, it was pointed out previously in \cite{Hegade:2022xij} without explicitly computing the scalar field solution in the exterior. 
Instead, by expanding the field equations near the horizon in Gaussian null coordinates and assuming {\it spherical symmetry}, the authors showed that in sGB gravity (and $F(R)$ gravity), regularity of the scalar field implies that the BH must be non-extremal. 
The contrapositive of this statement -- if the BH is extremal, the scalar field must be singular -- matches exactly what we find here. 
Although the theorem in~\cite{Hegade:2022xij} is not directly applicable to our setup, the underlying mechanism can be traced to the same origin, namely the monopolar part of the scalar field. 
As shown in the previous paragraph, in sGB gravity, when $s_{l = 0}(r) \neq 0$, the source generates a logarithmic term that diverges at the horizon. 
Meanwhile, in dCS gravity, the monopolar part of the source is \cite{McNees:2015srl, Lam:2025fzi}
\begin{equation}
    s_{l = 0}^{\rm dCS}(r) = \int_{-1}^{+1} \frac{48r \chi (r^2 - 3 \chi^2)(3r^2 - \chi^2)}{\Sigma^5} P_{l=0}(\chi) \,d\chi = 0, 
\end{equation}
which vanishes due to its odd parity\footnote{The full functional form of $s_l^{\rm dCS}(r)$ for dCS gravity can be found in Eq.~(29) of \cite{Lam:2025fzi}. }.
The leading-order term of the dCS pseudoscalar field then comes from the $l=1$ piece, whose near-horizon expansion yields 
\begin{equation}
    \vartheta_{l = 1}^{\rm dCS}(\rho) \sim \text{const} + \rho \log \rho + O(\rho).
\end{equation}
Any higher-order $l$-modes will generate terms that are proportional to $\rho^l \log \rho$.
These terms, despite not being infinitely differentiable, are finite at $\rho = 0$.
Thus, the dCS pseudoscalar field remains bounded at the horizon \cite{Horowitz:2026fzc}.

\subsection{Metric and Conserved Charges}

\begin{figure}[t!]
    \centering
    \includegraphics[width=\columnwidth]{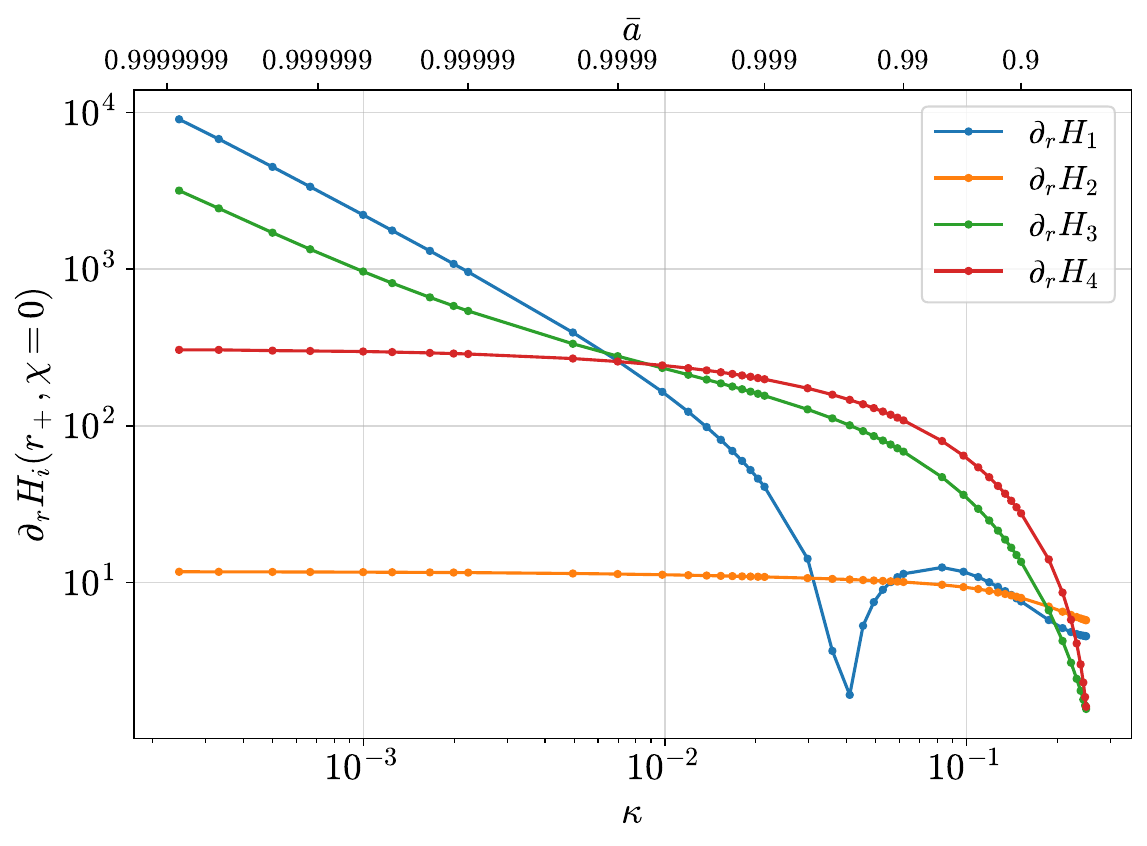}
    \caption{Identical to Fig.~\ref{fig:DrHi_kappa_dCS}, but for sGB gravity.}
    \label{fig:DrHi_kappa_sGB}
\end{figure}

The metric corrections at the horizon are shown in the left panel of Fig.~\ref{fig:Hi_kappa}. 
As $\kappa$ decreases, the $H_i(r_+, \chi)$ curves plateau, just as we observed in dCS gravity, despite $\vartheta_{\rm sGB}(r_+, \chi)$ diverging in this limit. 
As shown in the right panel of Fig.~\ref{fig:Hi_kappa}, the metric corrections also tend to finite values at spatial infinity. 
We find the same trend for any point that is away from the horizon and off the equatorial plane.   
In Fig.~\ref{fig:DrHi_kappa_sGB}, we show the radial derivative of the metric corrections $\partial_r H_i(r_+, \chi=0)$. 
As $\kappa$ becomes smaller, $\partial_r H_{2,4}$ flatten while $\partial_r H_{1,3}$ grow as $\kappa^{-1}$, which is much faster than that in the dCS case. 
We note that $\partial_r H_4$ flattens only on the equatorial plane; $\partial_r H_4(r_+, \chi \neq 0)$ goes as $\kappa^{-1}$ for small $\kappa$. 
Away from the horizon, $\partial_r H_i$ no longer increases as $\kappa^{-1}$; instead, the curves plateau. 

We also show the EFT corrections to the ADM mass $M^{(1)}$ in Fig.~\ref{fig:deltaM}, and the EFT corrections to ADM angular momentum $J^{(1)}$ in Fig.~\ref{fig:deltaJ}, which are computed using Eq.~\eqref{eq:ADM}. 
As in the dCS case, the sGB corrections are always positive.
Notice that the ADM mass correction first decreases with $\kappa$ up to $\sim 0.15$ ($\bar{a} \lesssim 0.9$). 
As $\kappa$ approaches zero, $M^{(1)}$ increases and approaches $\sim 3.0$ in the extremal limit. 
The $J^{(1)}$ correction increases monotonically and plateaus at small $\kappa$, with $J^{(1)} \to 8.9$.  

Let us examine the extremal limit of the metric corrections in more detail. 
Given that we observed a finite extremal limit, as in dCS gravity, the extremal metric correction is expected to admit the near-horizon expansion:
\begin{equation}
	H_i(r, \chi) = A_i(\chi) + B_i(\chi) \rho^{\beta} + \ldots, 
\end{equation}
for some $\beta > 0$ and some functions $A_i(\chi)$ and $B_i(\chi)$. 
Its radial derivative is therefore
\begin{equation}
	\partial_r H_i = B_i(\chi) \beta \rho^{\beta - 1} + \ldots.
\end{equation} 
As in dCS, we expect the extremal solution can be extrapolated to the case of small $\kappa$. 
Formally speaking, this is possible if the extremal expansion remains uniformly valid for the near-extremal solutions in a neighborhood of the horizon $\rho \lesssim O(\kappa)$. 
Suppose this is the case and, as the near-extremal horizon lies within the neighborhood, evaluating the above expression at $r=r_+$ gives
\begin{equation}
    \partial_r H_i(r_+,\chi) = O(\kappa^{\beta-1}) = o(\kappa^{-1}).
\end{equation}
However, the numerical results instead show
\begin{equation}
    \partial_r H_{i = 1,3}(r_+,\chi) \sim \kappa^{-1}, 
\end{equation}
and therefore, we have a contradiction. 
This implies that the assumption that the extremal expansion remains uniformly valid near the horizon cannot be true, and our expectation does not hold. 

\begin{figure}
    \centering
    \includegraphics[width=\columnwidth]{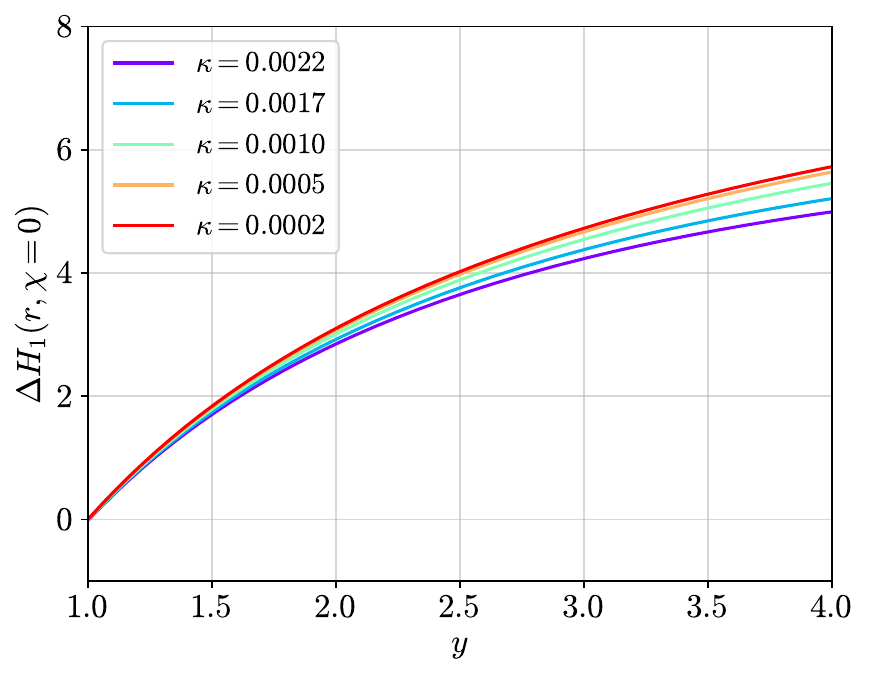}
    \caption{$\Delta H_1(r, \chi=0)$ against $y$, where $y = \rho/(2\kappa)$. Lines of different colors represent solutions with different $\kappa$. }
    \label{fig:H1_boundary_layer}
\end{figure}

This non-uniform behavior can be understood in terms of a boundary layer of width $O(\kappa)$ that develops just outside the near-extremal horizon.
To see this, as a representative example, we plot the near-extremal $\Delta H_1 \equiv H_1(r,\chi) - H_1(r_+,\chi)$ as a function of $y = \rho/(2\kappa)$ on the equatorial plane in Fig.~\ref{fig:H1_boundary_layer}. 
In this coordinate, the horizon is located at $y_+ = 1 + O(\kappa)$. 
In terms of $r$, the region $y_+ < y \lesssim 3/2$ corresponds to  $r_+ < r \lesssim r_+ + \kappa$. Observe that $dH_1/dy \sim O(1)$ at the horizon as $\kappa \to 0$.  This means that $H_1(r, \chi=0)$, along the radial direction, is changing on a scale of order $\kappa$ for arbitrarily small $\kappa$. In terms of derivatives with respect to $r$, this immediately reproduces the $\kappa^{-1}$ divergence we saw earlier: 
\begin{equation}
	\partial_r H_1(r_+, \chi = 0; \kappa) = \frac{1}{2\kappa} \frac{dH_1}{dy} \bigg|_{y = y_+}. 
\end{equation}
Similarly, we find the same boundary-layer behavior for $H_3$, while $H_4$ exhibits the corresponding $\kappa^{-1}$ derivative growth away from the equatorial plane.
This explains the limiting behavior of $H_i$ and $\partial_r H_i$. 

This boundary layer model also explicitly shows why the extremal limit is not smooth. 
Consider the near-horizon limit and the extremal limit of $y(r, \kappa)$:
\begin{align*}
	\lim_{\kappa \to 0^+} \lim_{r \to r_+} \frac{r - 1}{2\kappa} &= \lim_{\kappa \to 0^+} \left[1 + O(\kappa)\right] = 1, \\
	\lim_{r \to r_+} \lim_{\kappa \to 0^+} \frac{r - 1}{2\kappa} &\to \infty.
\end{align*}
Therefore, the two limits do not commute. 
Since $H_i$ depend on $y$ in the near-horizon regime, they do not converge uniformly in the neighborhood of the horizon.

As seen from Fig.~\ref{fig:Hi_kappa}, we expect the sGB solution to stay finite for any small $\kappa > 0$. 
However, the width of the boundary layer continues to shrink, and $\partial_r H_{i=1,3}(r_+, \chi)$ becomes large as $\kappa$ decreases. 
At fixed spectral order, resolving the boundary layer whose width shrinks with $\kappa$ eventually becomes prohibitively expensive.
Instead, one should use the rescaled coordinate $y$ as the radial coordinate in the boundary layer, and match the solution to an exterior solution with the correct boundary conditions.
Furthermore, since the extremal limit is non-uniform, the limiting solution obtained by taking $\kappa \to 0$ in Fig.~\ref{fig:Hi_kappa} is generally different from the extremal solution calculated by first setting $\kappa = 0$. 
This is already suggested in Appendix~\ref{sec:Lambda} by the near-horizon expansion of extremal sGB BHs, which generically exhibits a $\log\rho$ divergence in $H_i$ that is absent in the numerical limiting solution.
We will therefore refrain from constructing the full extremal solution numerically.

\subsection{The exterior-connected sGB near-horizon limit is not the sGB-deformed NHEK solution}

The same near-horizon expansion in Appendix~\ref{sec:Lambda} also clarifies why the sGB-deformed NHEK solution is not the throat of the asymptotically flat exterior constructed here.
Near the extremal horizon, the scalar field takes the schematic form
\begin{equation}
    \vartheta_{\rm sGB}(\rho,\chi)
    = c_{\log}\log\rho+\tilde{\vartheta}_0(\chi)+o(\rho^0).
\end{equation}
As shown in Appendix~\ref{sec:Lambda}, the leading angular equation admits two inequivalent branches.
If we require the scalar field to be regular at the poles $\chi=\pm1$, then $c_{\log}=4$ and the scalar field contains the unavoidable horizon term $4\log\rho$.
This is the branch selected by the asymptotically flat exterior.
If instead we set $c_{\log}=0$, the scalar field is regular at the horizon, but logarithmic singularities remain at the poles.
This second branch is the one obtained by solving the scalar equation directly on the NHEK background \cite{Chen:2018jed, Delgado:2020rev},
\begin{equation}\label{eq:NHEK_SF_sGB}
    \vartheta_{\text{sGB-NHEK}}(\chi)
    =
    2\log\left(\frac{1+\chi^2}{1-\chi^2}\right)
    -
    \frac{4(\chi^4+4\chi^2-1)}{(1+\chi^2)^3}.
\end{equation}

Thus, the two procedures impose different boundary conditions.
The exterior problem keeps the polar axis regular and accepts a logarithmic scalar divergence at the extremal horizon.
The local NHEK problem forces regularity at most points on the horizon, but it does so by making the poles singular. The two procedures therefore need not yield the same solution. 
Indeed, if we first solve the scalar equation in the full extremal Kerr exterior and then zoom into the near-horizon region, $\rho=\epsilon\tilde r$, the monopole term gives
\begin{equation}
    \vartheta_{l=0}
    =
    4\log\epsilon+4\log\tilde r+O(1).
\end{equation}
The constant $4\log\epsilon$ can be removed by the shift symmetry of the scalar field, but the $\log\tilde r$ term cannot.
The near-horizon limit of the exterior scalar field therefore retains radial dependence and does not reduce to an isometry-preserving NHEK profile.

The metric equations tell the same story.
Because the exterior-connected scalar contains $4\log\rho$, the extremal metric ansatz must be general enough to allow logarithmic terms,
\begin{equation}\label{eq:Hi_NearHorizonExpansion}
    H_i(r,\chi) = j_i(\chi)\log\rho+h_i(\chi)+O(\rho\log\rho),
\end{equation}
and the angular prefactor in Eq.~\eqref{eq:metric} must be generalized locally to
\begin{equation}\label{eq:modified_Xi}
    \Xi = 1 + \zeta\frac{\bar{M}^4}{r^4} \Big[\Upsilon(\chi) \log\rho + \Lambda\Big].
\end{equation}
Solving the near-horizon field equations with Eqs.~\eqref{eq:Hi_NearHorizonExpansion} and \eqref{eq:modified_Xi}, and requiring regularity at the poles fixes 
\begin{equation}\label{eq:Lambda_sGB_exterior}
    \Lambda_{\rm sGB} = -\frac{1097}{16\sqrt{2}},
\end{equation}
as derived in Appendix~\ref{sec:Lambda}.
If one instead uses the horizon-regular NHEK scalar field in Eq.~\eqref{eq:NHEK_SF_sGB} and a horizon-regular metric ansatz, one obtains\footnote{This value matches the sGB-deformed NHEK metric [see the value of $\delta\beta$ in Eq.~(37) of \cite{Chen:2018jed}], up to a factor of 64 due to a different convention of the coupling constant and metric ansatz.}
\begin{equation}\label{eq:Lambda_sGB_NHEK}
    \Lambda_{\rm sGB}^{\rm NHEK} = -\frac{969}{16\sqrt{2}}.
\end{equation}
The difference between Eqs.~\eqref{eq:Lambda_sGB_exterior} and \eqref{eq:Lambda_sGB_NHEK} reflects the fact that the sGB-deformed NHEK branch and the asymptotically flat exterior branch solve different local boundary-value problems.

The scalar divergence, together with the logarithmic metric terms allowed by the near-horizon equations, suggests that the linear EFT expansion we have been using may break down in this region.
A simple assumption about higher-order terms gives a clear picture of two different possibilities for the full extremal sGB solution. 
At zero coupling, there is a solution for any constant value of the scalar. 
In the near-horizon region, we can write this as $\vartheta \sim \rho^\gamma$ with $\gamma = 0$. 
When the higher derivative term is included with a small coefficient, $\gamma$ might be shifted to a small positive value. 
Similarly, metric corrections near the horizon might be given by $\rho^{\gamma_i}$ with $0< \gamma_i \ll 1$. 
(To first order in $\gamma_i$ these reproduce the $\log \rho$ terms we have seen.) 
If so, then the full solution would still have a continuous extremal limit. 
Alternatively, if $\gamma_i < 0$, the metric might blow up at the horizon, producing a strong singularity. 

We therefore conclude that the sGB-deformed NHEK solution solves a different local near-horizon boundary-value problem and is not the near-horizon limit of the regular-pole, asymptotically flat exterior constructed here. 
At linear order, with the $\log \rho$ terms given in Eq.~\eqref{eq:Hi_NearHorizonExpansion}, although the proper radial distance to the extremal horizon (on a stationary surface) diverges and shows a long near-horizon region as $\kappa \to 0$, the geometry is no longer the same as NHEK. 
Furthermore, the fact that the scalar field and/or metric corrections near an extremal horizon become large suggests the linear approximation we have been using breaks down. So we will refrain from examining the structure of this new infinite throat.

\section{Breakdown of Effective Field Theory}\label{sec:BreakdownEFT}

We now return to the near-extremal numerical solutions.
Knowing that in both the dCS and sGB solutions, radial derivatives become large when $\kappa$ is small, one would expect certain curvature scalars and observables to become large as well, causing the breakdown of the EFT. 
In this section, we compute the Kretschmann and tidal forces, showing that these quantities can become large for near-extremal BHs in dCS and sGB gravity, violating the perturbative nature of EFTs. 

\subsection{Classification}

First, we elaborate on the breakdown of the EFT description. 
As briefly discussed in Sec.~\ref{sec:introduction}, the EFT ceases to be well controlled once the leading-order correction becomes comparable to the corresponding zeroth-order quantity. 
Broadly speaking, this breakdown can be classified according to which perturbative expansion fails:
\begin{enumerate}
    \item The EFT expansion of the action becomes uncontrolled;
    \item The perturbative expansions of observables become uncontrolled.
\end{enumerate}
The first category arises when the higher-derivative terms in the EFT become arbitrarily large.  
In many higher-derivative EFTs, the Lagrangian is constructed by combinations of scalar curvature invariants \cite{dCS_01, dCS_02, Yagi:2015oca, Nair:2019iur, Cano_Ruiperez_2019, Bueno:2016xff}. 
Near regions where scalar curvature invariants diverge, i.e., near scalar polynomial curvature singularities \cite{Hawking:1973uf}, higher-derivative terms become large, and the on-shell Lagrangian diverges. 
Consequently, the perturbative expansion of the EFT effective action could break down, and the variation of the action around the background solution becomes ill-defined. 

The second category can be associated with the recently found tidal force singularities \cite{Horowitz:2023xyl, Horowitz:2024dch, Horowitz:2024kcx,Horowitz:2026fzc} and QNM amplification \cite{Cano:2025ejw, DiRusso:2025qpf, Boyce:2025fpr, Husken:2026axq, Cano:2025mht}. 
As corrections to observables become large, the corresponding perturbative series will likely become uncontrolled and the EFT expansion will break down.
In particular, the tidal force singularity also indicates the breakdown of worldline EFT, because the (contracted) Weyl tensor becomes divergent \cite{Horowitz:2024dch, Chen:2024sgx}. 
The worldline EFT effective action, written as a series in (contracted) Weyl tensors, becomes ill-defined. 
This type of breakdown of EFT was further explored by \cite{Chen:2024sgx}, clarifying its relationship with parallel-propagated curvature singularities \cite{Hawking:1973uf}.

\subsection{Scalar Curvature Invariants}

\begin{figure}[t!]
    \centering
    \includegraphics[width=\columnwidth]{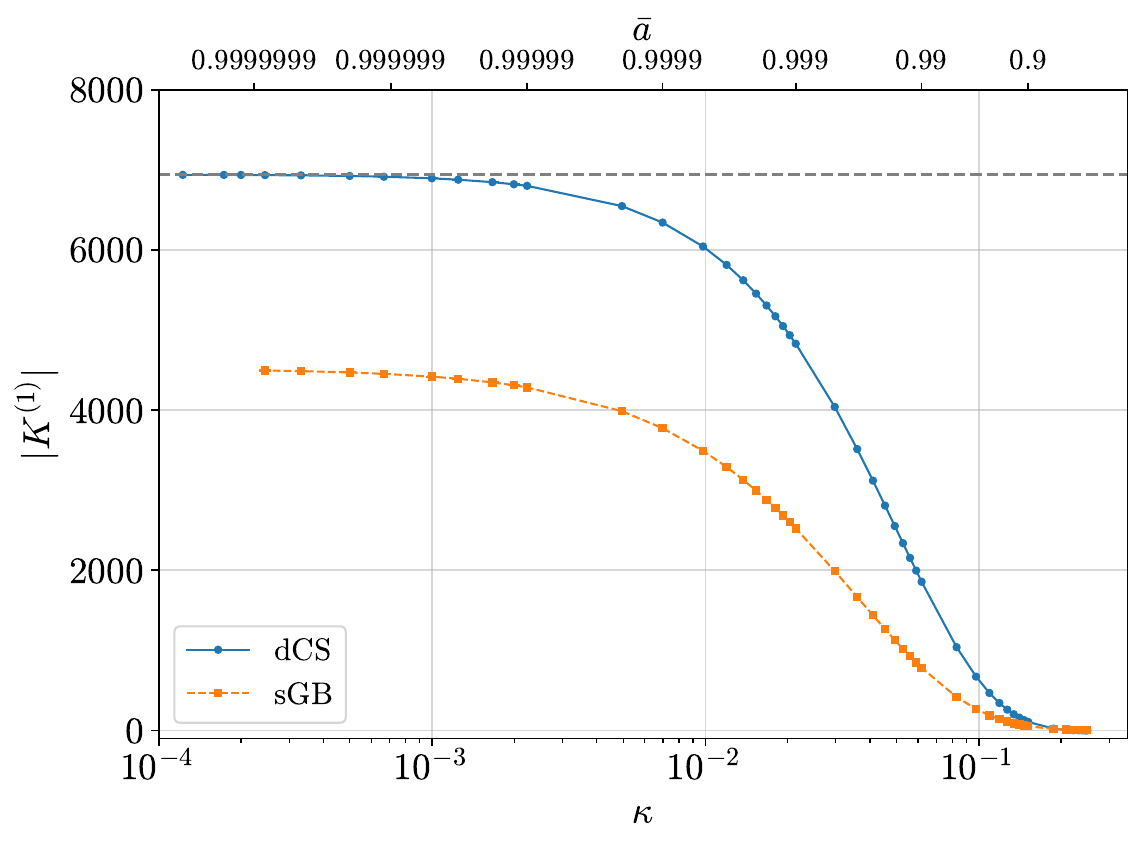}
    \caption{The EFT correction $|K^{(1)}|$ to the Kretschmann scalar in dCS and sGB gravity and in $\bar M=1$ units, evaluated on the equatorial plane at $r_+$. Horizontal dashed lines represent the Kretschmann correction computed using the dCS extremal solution. }
    \label{fig:Kretschmann}
\end{figure}

With the breakdown of EFT explained, we will now investigate whether dCS and sGB break down in the near-extremal regime. 
A natural geometric quantity to study is the Kretschmann curvature scalar $K \equiv R_{\mu\nu\alpha\beta} R^{\mu\nu\alpha\beta}$, as it is commonly used to characterize the curvature of spacetime. 
In Kerr, the Kretschmann is 
\begin{equation}
    K^{(0)}(r,\chi) = \frac{48\left(r^6 - 15r^4\bar{a}^2\chi^2 + 15r^2\bar{a}^4\chi^4 - \bar{a}^6\chi^6\right)}{\Sigma^6}\,,
\end{equation}
and in particular, the equatorial Kerr value is
\begin{equation}
    K^{(0)}(r,0)=\frac{48}{r^6},
\end{equation}
so $K^{(0)}(r_+,0)=48/r_+^6$, where recall we are working in units in which $\bar M=1$.

Figure~\ref{fig:Kretschmann} shows the EFT correction to the Kretschmann scalar on the equator and on the horizon $|K^{(1)}(r_+, 0)|$ as a function of $\kappa$.
In both theories, as $\kappa \to 0$, $|K^{(1)}|$ at the horizon increases rapidly at first, and then it plateaus. 
Observe that the correction becomes two orders of magnitude larger than the Kerr Kretschmann $K^{(0)}$. 
Despite $K^{(1)}$ containing $\partial_r H_i$, the Kretschmann corrections at the horizon plateau as $\kappa$ becomes small. 
This is not a coincidence. 
For any scalar curvature invariant, any two derivatives must be supplemented by an inverse metric $g^{\mu\nu}$ \cite{Horowitz:2024dch}. 
Near extremality, $g^{rr} \sim 4\kappa(r - r_+)$ close to the horizon. 
This cancels out the $\log\kappa$ and $\kappa^{-1}$ growth in dCS and sGB gravity, respectively.
Therefore, the extremal dCS Kretschmann does not diverge, and we expect the sGB Kretschmann curve to stay flat for even smaller $\kappa$.

In dCS gravity, since both the scalar field and Kretschmann are finite at extremality, the action does not diverge. 
However, in sGB gravity, the on-shell Lagrangian diverges at the horizon as $\vartheta_{\rm sGB}$ diverges, so the quadratic coupling term also blows up. 
One might wonder if the action would diverge as well. 
To compute the action on-shell, we utilize the exact solution in Appendix~\ref{sec:sGB_Scalar_Field}. 
The quadratic coupling term at leading order in $\zeta$ can be written as\footnote{$K^{(1)}$ does not enter the quadratic coupling term to leading order in $\zeta$. }
\begin{equation}
    \alpha \varphi \mathscr{Q} = -\frac{\zeta}{\Sigma} \left[\sum_{l=0}^{\infty} \vartheta_l(r) P_l(\chi)\right] \left[\sum_{l=0}^{\infty} s_l(r) P_l(\chi)\right].
\end{equation}
Integrating over the exterior, we have
\begin{equation}
    \int \sqrt{-g} d^4x \alpha \varphi \mathscr{Q} = -\zeta \sum_{l=0}^{\infty} \frac{2}{2l+1} \int dt d\phi \int_{r_+}^{\infty} dr \, \vartheta_l(r) s_l(r). 
\end{equation}
At the horizon, since $\vartheta_l(r)$ is at most logarithmically divergent while $s_l(r)$ is finite, the integral over $r$ converges and the action is finite. 
Thus, the quadratic coupling is singular at the extremal horizon, but the singularity is only logarithmic and is integrable. 
The on-shell action is therefore finite, and in this restricted sense sGB gravity, to leading-order in $\zeta$, does not exhibit the first type of EFT breakdown.

\subsection{Divergence of Tidal Forces}

Motivated by the tidal force singularities observed in other higher-derivative theories, let us now investigate the tidal force at the horizon. 
Consider the geodesic deviation equation satisfied by the deviation vector $\xi^{\mu}$ between neighboring ingoing null congruences $l^{\mu}$ \cite{Poisson:2009pwt, Wald:1984cw}: 
\begin{equation}
    \frac{d^2 \xi^{\mu}}{d\lambda^2} = -R^{\mu}{}_{\alpha \nu \beta} l^{\alpha} \xi^{\nu} l^{\beta}. 
\end{equation}
We define the tidal tensor
\begin{equation}\label{eq:tidaltensor}
    {\cal R}_{\mu\nu} = R_{\mu\alpha\nu\beta} l^{\alpha} l^{\beta}
\end{equation}
which represents the geodesic deviation along $l^{\mu}$. 
Since reflection symmetry is manifest, equatorial geodesics remain contained in the equatorial plane \cite{Cano_Ruiperez_2019, Lam:2025fzi}. 
Henceforth, we focus on zero angular momentum, ingoing, null congruences on the equatorial plane. 
Following \cite{Horowitz:2024dch}, we find
\begin{equation}\label{eq:nullgeod}
\begin{split}
    l^{\mu} &= 
    \left.
    \begin{bmatrix}
    \frac{\cal A}{r^2 \Delta}(1 - \zeta H_1) \\
    -\frac{\sqrt{\cal A}}{r^2}\left[1 - \frac{\zeta}{2}(H_1 + H_3)\right] \\
    0 \\
    \frac{2\bar{a}}{r\Delta}\left[1 + \zeta (H_2 - H_1)\right]
    \end{bmatrix} \right|_{\chi = 0}
\end{split}
\end{equation}
where $H_i$ and $\cal A$ are evaluated at $\chi = 0$.  We fix the normalization of the ingoing null vector by choosing unit energy $l_t = g_{t\mu}l^\mu = -1$.
We note that $l^{\mu}$ appears to diverge in the limit as $r \to r_+$; however, this is a coordinate artifact that arises from using Boyer-Lindquist coordinates. 
This divergence can be eliminated by using ingoing Eddington-Finkelstein coordinates or any horizon-penetrating coordinates. 
These coordinate singularities disappear in the calculation of the tidal tensor once the limit is taken appropriately, as shown later in this subsection.

\begin{figure}[t!]
    \centering
    \includegraphics[width=\columnwidth]{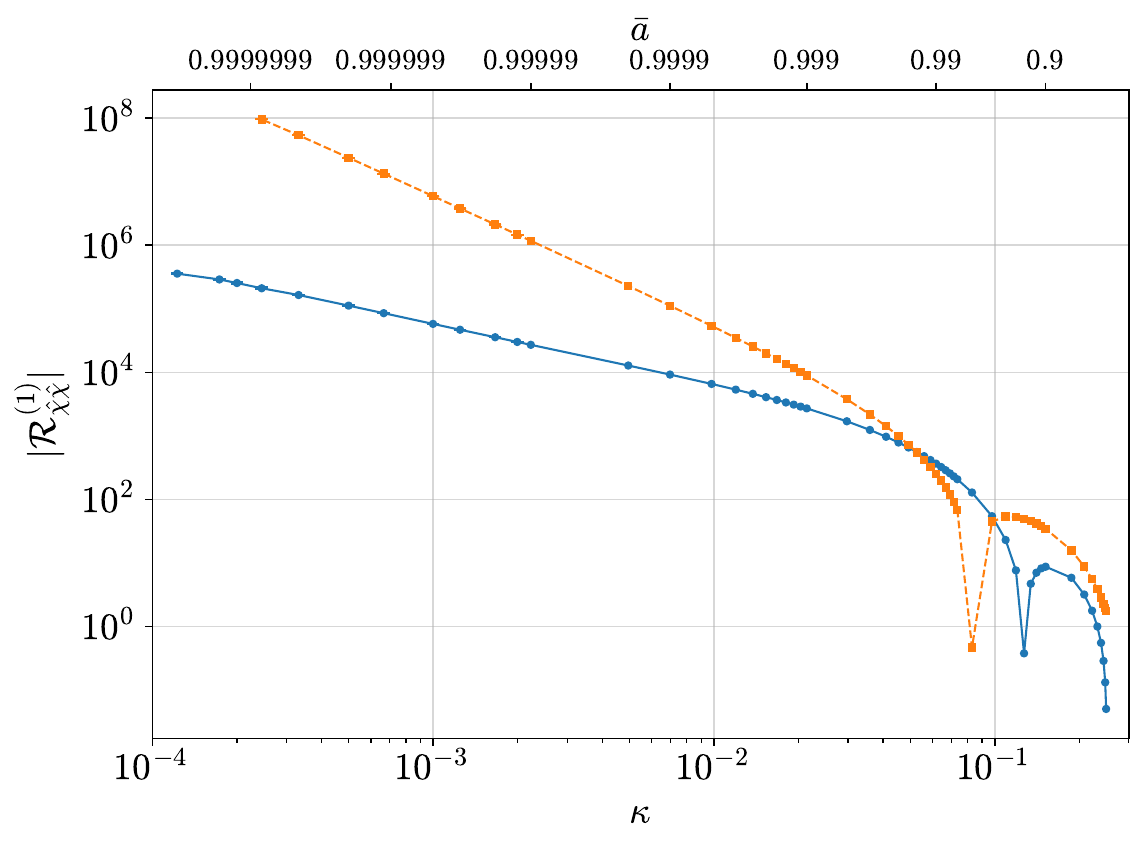}
    \caption{The orthonormal polar tidal force $|{\cal R}_{\hat\chi\hat\chi}^{(1)}|$ in dCS and sGB gravity, in $\bar M=1$ units. Near $\kappa = 0.08$ and $0.12$, ${\cal R}_{\hat\chi\hat\chi}^{(1)}$ have zero-crossings. For small $\kappa$, the tidal force in dCS diverges as $\kappa^{-1}$, while it diverges as $\kappa^{-2}$ in sGB. }
    \label{fig:Rchichi}
\end{figure}

We now choose the deviation vector to point in the polar direction. 
More precisely, we write\footnote{Since geodesic deviation describes the relative acceleration of \textit{infinitesimally} nearby geodesics, we can still use Eq. \eqref{eq:nullgeod} for the null tangent vectors as reference vectors on the equatorial plane.}
\begin{equation}
    \xi^\mu = \hat e_\chi^\mu ,
\end{equation}
where $\hat e_\chi^\mu$ is the unit vector in the $\chi$ direction, normalized with the metric in Eq.~\eqref{eq:metric}. 
On the equatorial plane, this vector is
\begin{equation}
    \hat e_\chi^\mu
    =
    \frac{(\partial_\chi)^\mu}{\sqrt{g_{\chi\chi}}}\bigg|_{\chi=0}.
\end{equation}
We then define the polar tidal force along the null congruence as
\begin{equation}
   {\cal R}_{\hat\chi\hat\chi}
   \equiv
   {\cal R}_{\mu\nu}\hat e_\chi^\mu \hat e_\chi^\nu
   =
   R_{\mu\alpha\nu\beta}\hat e_\chi^\mu l^\alpha \hat e_\chi^\nu l^\beta .
\end{equation}

We pick the $\chi\chi$-component because this gives us the largest contribution of the tidal force on the horizon. 
To evaluate this at the horizon, one must be careful about spurious divergences due to numerical error.  
The tidal force correction generally can be written as a linear combination of $H_i$ and its derivatives, 
\begin{equation}
    {\cal R}_{\hat\chi\hat\chi}(r) = \sum_{i=1}^{4} \sum_{j,k} \alpha_{ijk}(r) \partial_r^j \partial_{\chi}^k H_i(r,0). 
\end{equation}
As $l^{\mu}$ is proportional to inverse powers of $\Delta$, $\alpha_{ijk}(r)$ generally diverges at the horizon. 
If we expand $\alpha_{ijk}(r)$ near the horizon, we find
\begin{align}
    {\cal R}_{\hat\chi\hat\chi}(r) &= \sum_{p=-1}^{\infty} \left[\sum_{i=1}^{4} \sum_{j,k} \tilde{\alpha}_{ijk,p} \partial_r^j \partial_{\chi}^k H_i(r,0)\right] (r - r_+)^p 
    \nonumber \\
    &= \sum_{p=-1}^{\infty} {\cal R}_{\hat\chi\hat\chi, p}(r) (r - r_+)^p, 
\end{align}
where the $\tilde{\alpha}_{ijk,p}$ are constants. 
We do not expand $\partial_r^j \partial_{\chi}^k H_i$ near the horizon since they could mimic non-analytic terms that cannot be Taylor expanded. 
Notice the summation index $p$ starts at $-1$, thus ${\cal R}_{\hat\chi\hat\chi}$ diverges at the horizon unless ${\cal R}_{\hat\chi\hat\chi, p=-1}(r_+)$ vanishes. 

The $p=-1$ pole is not a physical contribution to the tidal force. 
It is proportional to the angular dependence of the surface gravity. 
Explicitly, 
\begin{equation}
    {\cal R}_{\hat\chi\hat\chi, p=-1}(r) \sim \left\{\frac{\partial^2}{\partial\chi^2}\Big[H_1(r, \chi) - H_3(r, \chi)\Big]\right\}_{\chi=0}.
\end{equation}
At the horizon, this quantity should vanish exactly according to the properties of the Killing horizon \cite{Racz:1992bp}. 
Or equivalently, by the boundary condition [Eq.~\eqref{eq:HorizonBC}] imposed. 
In the numerical solution, however, small residual violations of this boundary condition can be amplified by the explicit $(r-r_+)^{-1}$ factor, so we remove this pole by hand before taking the horizon limit. 

The finite tidal force is then
\begin{equation}
    {\cal R}_{\hat\chi\hat\chi}(r_+) = {\cal R}_{\hat\chi\hat\chi,p=0}(r_+),
\end{equation}
which in Kerr reduces to 
\begin{equation}
    {\cal R}_{\hat\chi\hat\chi}^{(0)}(r_+) = \frac{3\bar{a}^2}{r_+^5}. 
\end{equation}
Figure~\ref{fig:Rchichi} shows $|{\cal R}_{\hat\chi\hat\chi}^{(1)}|$ for dCS and sGB gravity. 
When $\kappa < 10^{-2}$, it is evident that the tidal forces diverge as $\kappa^k$. 
Fitting over that region, we find that in dCS, $k \approx -0.93$, while in sGB $k \approx -2.0$. 
One might worry that such divergences are gauge or tetrad artifacts. 
In Appendix~\ref{sec:R2}, we show the correction to the square of the tidal tensor $|({\cal R}_{\mu\nu}{\cal R}^{\mu\nu})^{(1)}|$, whose behavior is similar to $|{\cal R}_{\hat\chi\hat\chi}^{(1)}|$.
Although the leading-order on-shell action remains finite in the extremal limit, the tidal force felt by an infalling observer diverges.
Although the calculation above was performed for null congruences, we have checked that the same tidal-force blow-up occurs for infalling timelike observers.
Therefore, both dCS and sGB gravity suffer from EFT breakdown of the second type.

\subsection{Regime of Validity}

\begin{figure}[t!]
    \centering
    \includegraphics[width=\columnwidth]{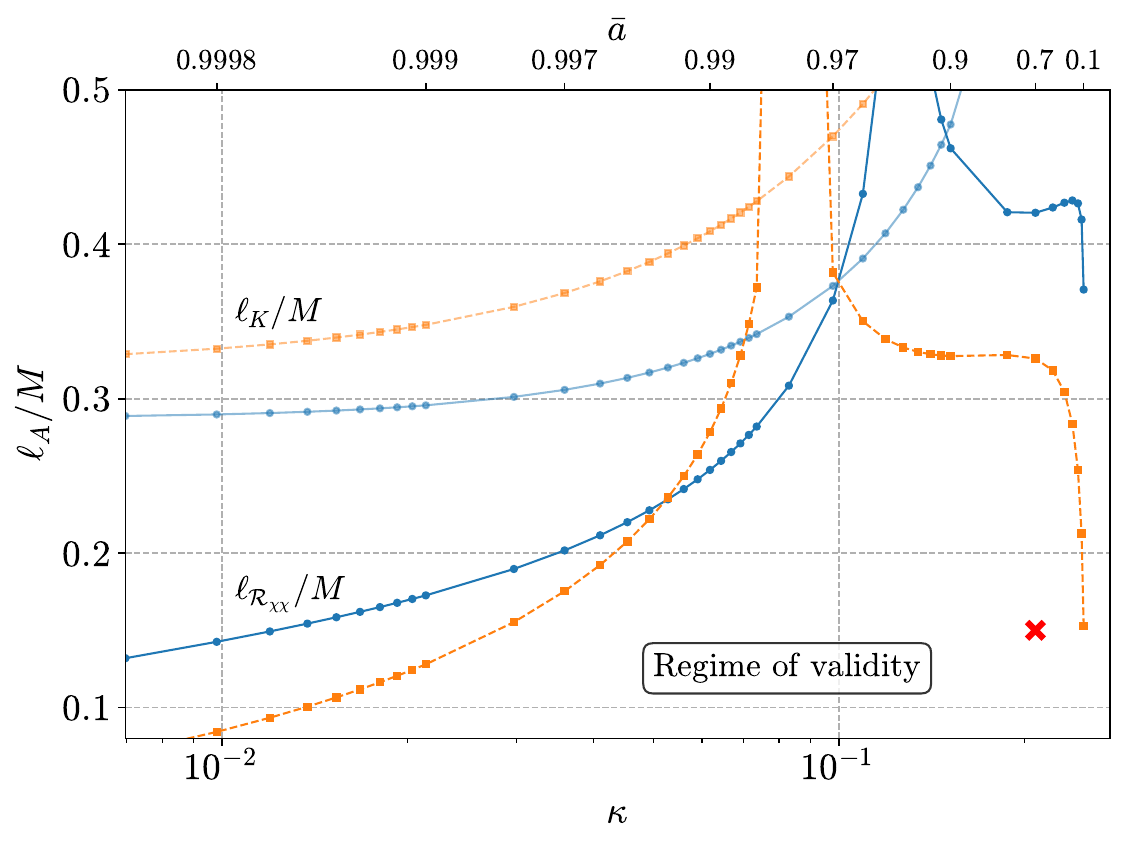}
    \caption{The regime of validity (region under the curves) of dCS and sGB gravity, defined by $\ell_{{\cal R}_{\hat\chi\hat\chi}}/M$ and $\ell_K/M$. Blue solid lines represent the dCS couplings, and orange dashed lines show the sGB couplings. The top translucent lines display $\ell_K/M$ while the lower opaque lines show $\ell_{{\cal R}_{\hat\chi\hat\chi}}/M$. 
    The red cross represents current observational constraints on dCS gravity given in \cite{ringdown_test_01}. } 
    \label{fig:Regime_of_Validity}
\end{figure}

Since the tidal force diverges, the EFT becomes uncontrolled in the extremal limit. 
For near-extremal BHs, although the tidal force correction does not diverge, it becomes sufficiently large that, despite it being suppressed by the small coupling constant $\zeta$, the correction $\zeta {\cal R}_{\hat\chi\hat\chi}^{(1)}$ could overwhelm the original Kerr tidal force ${\cal R}_{\hat\chi\hat\chi}^{(0)}$. 
In that regime, the leading-order perturbative expansion is no longer reliable.
One way to estimate the validity of the EFT expansion is to see whether the first-order EFT correction becomes larger than the GR quantity, i.e., for any $A$, $\zeta |A^{(1)}/A^{(0)}|$ should be less than unity, so that the perturbative expansion is under control\footnote{At points where $A^{(0)} = 0$ or $A^{(1)} = 0$, this condition is no longer well-defined. In those cases, one would need to investigate higher-order terms, such as $A^{(2)}$ in order to estimate the regime of validity of the perturbative series.}. 
We can then compute the leading-order estimate of the regime of validity in both theories and place an upper bound on the coupling constant $\zeta_{A} \equiv |A^{(0)}/A^{(1)}|$.
Equivalently, we can compute the upper bound of the characteristic length scale 
\begin{equation}
    \frac{\ell_A}{M} = \zeta_A^{1/4}, 
\end{equation}
which would allow us to compare with existing observational constraints easily. Importantly, these ``upper bounds'' are not actual constraints; rather, they tell us about the mathematical boundary of the regime of validity of the EFT approximation (without the use of experimental data or observations). 

First, let us illustrate the idea with the tidal force and the Kretschmann at the horizon, because both quantities become large in the near-extremal regime, as compared to their Kerr values. 
Figure~\ref{fig:Regime_of_Validity} shows $\ell_{{\cal R}_{\hat\chi\hat\chi}}/M$ and $\ell_K/M$ as a function of $\kappa$.
The top translucent lines represent the dCS and sGB gravity $\ell_K$, while the bottom ones show $\ell_{{\cal R}_{\hat\chi\hat\chi}}$. 
Observe that, as $\kappa$ decreases, the $\ell_K$ decrease monotonically.
On the other hand, at small $\bar{a}$, $\ell_{{\cal R}_{\hat\chi\hat\chi}}$ is small because ${\cal R}_{\hat\chi\hat\chi}^{(0)} \sim \bar{a}^2$. 
As $\kappa$ decreases, $\ell_{{\cal R}_{\hat\chi\hat\chi}}$ increases in both theories and then diverges.
This is because ${\cal R}_{\hat\chi\hat\chi}^{(1)}$ crosses zero at $\kappa \approx 0.08$ in sGB and $\kappa \approx 0.12$ in dCS, as also seen in Fig.~\ref{fig:Rchichi}. 
After that zero crossing, $\ell_{{\cal R}_{\hat\chi\hat\chi}}$ decreases uniformly.
The dCS maximum coupling declines roughly as $\kappa^{1/4}$, while the sGB one decreases with $\kappa^{1/2}$. 
This is consistent with the divergent behavior seen in Fig.~\ref{fig:Rchichi}. We then expect that, at extremality, $\ell_{{\cal R}_{\hat\chi\hat\chi}} = 0$.

We now combine the Kretschmann and tidal-force diagnostics by taking the lower envelope of the individual bounds,
\begin{equation}\label{eq:zeta_max}
    \ell_{\rm max} = \min_A \ell_A .
\end{equation}
Of course, one cannot check every possible diagnostic.
One could construct arbitrarily high scalar curvature invariants, or consider other quantities such as the ADM mass and angular momentum.
The point, however, is that the near-extremal regime is controlled by the fastest-growing correction.
For the solutions studied here, the scalar invariants we have checked and the ADM corrections remain finite.
Thus, at small $\kappa$, the tidal force sets the dominant restriction.
Away from this limit, other diagnostics could in principle matter.
As a check, we have also evaluated the cubic invariant
\begin{equation}
    I_3
    =
    R_{\mu\nu}{}^{\alpha\beta}
    R_{\alpha\beta}{}^{\rho\sigma}
    R_{\rho\sigma}{}^{\mu\nu},
\end{equation}
as well as the ADM quantities, and we find that including them changes $\ell_{\rm max}$ only mildly.

The combined regime of validity of dCS and sGB gravity is the region below the lower envelope of the $\ell_K$ and $\ell_{{\cal R}_{\hat\chi\hat\chi}}$ curves in Fig.~\ref{fig:Regime_of_Validity}.
For $\kappa > 0.06$, the maximum allowed coupling is $\ell_{\rm max} \approx 0.4 M$ in dCS gravity and $\ell_{\rm max} \approx 0.3 M$ in sGB gravity.
For $\kappa < 0.07$, $\ell_{\rm max}$ is determined by $\ell_{{\cal R}_{\hat\chi\hat\chi}}$, because the tidal force grows fastest and dominates the other $\ell_A$ bounds.
Compared with current experimental bounds, the typical constraints from gravitational wave BH observations lie within the regime of validity. 
For example, from the combined inspiral-merger-ringdown analyses of two high signal-to-noise ratio events, GW150914 and GW200129, Ref.~\cite{ringdown_test_01} found that $\ell < 13.7 \,{\rm km}$ for dCS gravity\footnote{Due to the difference in the Lagrangian density, the characteristic length scale $\ell_S$ defined in \cite{ringdown_test_01} is $2\sqrt{2}$ times ours, that is, $\ell_S = 2\sqrt{2} \ell$. }. 
The source frame remnant masses of these two events are roughly $60\,M_{\odot}$ and have dimensionless spin $a \approx 0.7$, leading to an upper bound of $\ell / M \approx 0.15$. 
We plot the corresponding point in Fig.~\ref{fig:Regime_of_Validity} with a red cross, which clearly shows that the inferred bound is consistent with the EFT approach. 
Similarly, in sGB gravity, by stacking prominent events in the GWTC-1 and GWTC-2 catalogs, Ref.~\cite{Perkins:2021mhb} found that $\ell < 3.2\,{\rm km}$. 
Using the same remnant mass, this translates to $\ell / M = 0.036$, which is hidden below the $y$-axis of Fig.~\ref{fig:Regime_of_Validity}. 
This confirms that the constraints to date are consistent with the EFT assumptions.

\section{Conclusion}\label{sec:Conclusion}

In this work, we have computed stationary, axisymmetric (near-)extremal BH solutions in dCS and sGB gravity, which are leading-order EFTs of gravity that contain one dynamical pseudoscalar or scalar field. 
Here, we extend our previous calculations to surface gravities as small as $\kappa \sim 10^{-4}$, corresponding to rapidly rotating solutions with $1 - \bar{a} \sim O(10^{-8})$.
In dCS gravity, the same horizon-adapted framework can be continued directly to zero surface gravity, yielding an extremal BH solution whose metric corrections connect continuously to the sub-extremal family.
The extremal solution's near-horizon limit matches the dCS-deformed NHEK solution found previously in \cite{Chen:2018jed, Cano:2023dyg}. Thus, the solution still has the same throat structure and enhanced symmetries as Kerr extremal BHs.
In sGB gravity, as $\kappa \to 0$, the scalar field develops an unavoidable logarithmic divergence at the horizon, while the exterior metric develops a boundary layer of width $\kappa$ near the horizon; the extremal limit is therefore nonuniform, and the near-extremal metric corrections are generally not connected to the extremal ones. 
At extremality, we find that the sGB-deformed NHEK solution solves a different local boundary-value problem than the full exterior solution: it is not the near-horizon limit of the regular-pole, asymptotically flat exterior branch constructed here.

These EFT-corrected BHs develop large curvature corrections and tidal forces at the horizon. 
Close to the extremal limit, we found that the correction to the Kretschmann (with the coupling factored out) becomes two orders of magnitude larger than that in Kerr, while the tidal force grows with inverse powers of the surface gravity. 
Perturbative self-consistency then requires the coupling to be sufficiently small so that these corrections remain small deformations of the GR result.
Based on this argument, we estimate the regime of validity of dCS and sGB gravity by requiring the Kretschmann and tidal force corrections to be less than the Kerr ones. 
We find that current observational bounds lie within the regime of validity estimated here, so present BH constraints do not by themselves push either theory outside perturbative EFT control.

A direct application of our (near-)extremal spectral solution is to compute quasinormal modes (QNMs) of (near-)extremal BHs. 
As done in \cite{Husken:2026axq, Cano:2025ejw}, we can study the phase boundary between zero-damped modes and damped modes to understand the amplification of QNMs. 
The linear stability of dCS- and sGB-corrected BHs can also be probed by looking at whether these modes become unstable \cite{Husken:2026axq, Boyce:2025fpr, Boyce:2026rnn}.
The QNM frequencies can be computed with, for instance, the METRICS approach \cite{Chung:2023zdq, Chung:2023wkd, Chung:2024vaf, Chung:2024ira, Chung:2025gyg}, the modified Teukolsky formalism \cite{Li:2022pcy, Li:2025fci, Wagle:2023fwl}, or potentially the continued fractions method \cite{Leaver_01, Karikos:2026arz}. 
Moreover, because the dCS-deformed NHEK solution is connected to the asymptotically flat exterior, one can use the throat geometry to study near-horizon mode dynamics, as in~\cite{Cano:2024bhh}.

Several other extensions of this work could also be explored. 
One natural direction is to characterize the near-horizon scaling data of the extremal solutions.
In dCS gravity, this information can be extracted from the $(r-r_+)\log(r-r_+)$ behavior of the metric and interpreted in terms of the scaling dimensions of the dCS-deformed NHEK throat.
These near-horizon scaling data control the tidal divergence found here and should also determine how the Aretakis instability is modified at the extremal horizon \cite{Aretakis:2011ha,Aretakis:2012ei,Lucietti:2012sf,Angelopoulos:2018yvt,Chen:2025sim}.
As the scaling dimension is modified, the Aretakis instability could either be tamed or enhanced, depending on the sign of the modification \cite{Chen:2025sim}. 
In sGB gravity, due to the leading $\log\rho$ terms in the near-horizon extremal solution, this suggests the linear EFT expansion breaks down. 
Higher-order terms in the action coming from different UV completions should be considered to see whether higher-order corrections could regularize the logarithmic divergence.

One could also extend and implement our framework to other modified theories of gravity, such as dCS and sGB gravity with different coupling functions \cite{Kleihaus:2011tg, Kleihaus:2015aje} or with massive degrees of freedom \cite{Chung:2026zon, Richards:2023xsr}. 
For sGB gravity with quadratic or exponential couplings, previous numerical studies suggest that the extremal limit would remain regular; it would be interesting to see whether the tidal force singularities and the enhanced NHEK geometry are still there. 
For theories with massive fields, recent work has shown that these fields could introduce stronger tidal force singularities \cite{Horowitz:2026fzc}.
In the NHEK spacetime, massive fields are subject to the Breitenlohner–Freedman bound, thus providing an effective mass bound to the scalar field \cite{Durkee:2010ea, Dias:2012pp, Chen:2024sgx}. 
We leave these extensions to future work.

\section*{Acknowledgment}

We thank Leo Stein and Jorge Santos for comments and suggestions on the draft. 
K.~K.~H.~L. and N.~Y. acknowledge support from the Simons Foundation through Award No. 896696, the Simons Foundation International under grant SFI-MPS-BH-00012593-01, and the NSF through Grant No.~PHY 25-12423. G.~H. was supported in part by NSF grant PHY-2408110 and by Simons Foundation International and the Simons Foundation through Simons Foundation grant SFI-MPS-BH-00012593-08. 
Some calculations and results reported in this paper were produced using the computational resources of the Illinois Campus Cluster, a computing resource that is operated by the Illinois Campus Cluster Program (ICCP) in conjunction with National Center for Supercomputing Applications (NCSA), and is supported by funds from the University of Illinois at Urbana-Champaign.

\onecolumngrid
\appendix

\section{Near-Horizon Expansion and Metric-Ansatz Constant}\label{sec:Lambda}
\subsection{\MakeLowercase{d}CS gravity}

In dCS gravity, at the extremal limit with $\bar{a} = 1$ (recall we set $\bar{M} = 1$), assuming the horizon is finite, we can expand the metric functions as 
\begin{equation}
    H_i(r, \chi) = h_i(\chi) + k_i(\chi) \rho \log \rho + p_i(\chi) \rho + o(\rho),  
\end{equation}
where recall that $\rho = r - 1$, which was defined below Eq.~\eqref{eq:d-drho}. 
Substituting the above expansion and the scalar field solution in Eq.~\eqref{eq:NHEK_SF_dCS}, or the full exterior solution given in \cite{McNees:2015srl}, into the field equations, we find that in dCS gravity, 
\begin{equation}
    \frac{d}{d\chi}\left[(1 - \chi^2)^{3/2} \frac{d}{d\chi}(h_1 - h_4)\right] = 2\Lambda \sqrt{1 - \chi^2} + T(\chi), 
\end{equation}
where $T(\chi)$ is a function that contains the source terms of the field equations. 
Integrating both sides from $-1$ to $+1$, the left-hand-side vanishes if $h_1 - h_4$ is regular. Therefore, this forces a constraint on $\Lambda$ where the integral of the right-hand side must be zero. Then, we find 
\begin{equation}
    \Lambda = -\frac{975}{16\sqrt{2}}.
\end{equation}
If one proceeds to solve the remaining dCS near-horizon field equations, one finds the dCS-deformed NHEK solution presented in Eq.~\eqref{eq:dCS_NHEK_sol}, with an additional integration constant that can be fixed by requiring the absence of conical singularities, $h_3(\chi = \pm 1) = h_4(\chi = \pm 1)$. 

\subsection{\MakeLowercase{s}GB gravity}

In sGB gravity, we wish to compute the near-horizon extremal solution.
First, we must compute the near-horizon scalar field solution. 
The scalar field equation in sGB gravity can be written as 
\begin{equation}
    \Box \vartheta = -R^{\mu\nu\alpha\beta}R_{\mu\nu\alpha\beta}\,,
\end{equation}
which, after multiplying by $\Sigma$, reduces to 
\begin{equation}
    \frac{\partial}{\partial r} \left[(r - 1)^2 \frac{\partial \vartheta}{\partial r}\right] + \frac{\partial}{\partial \chi} \left[(1 - \chi^2)\frac{\partial \vartheta}{\partial \chi}\right] = - \frac{48 (r^6 - 15 r^4 \chi^2 + 15 r^2 \chi^4 - \chi^6)}{\Sigma^5}, 
\end{equation}
when evaluated on the extremal Kerr background.
We expand the scalar field as
\begin{equation}
    \vartheta(\rho, \chi) = \sum_{n = 0}^{\infty} \tilde{\vartheta}_n(\chi) \rho^n + \sum_{n = 0}^{\infty} \bar{\vartheta}_n(\chi) \rho^n \log\rho\,,
\end{equation}
and solve the scalar field equation order by order in $\rho$. First, at order $\log\rho$, the source vanishes and regularity implies $\bar{\vartheta}_0(\chi)=c_{\log}$.
At order $\rho^0$, we find
\begin{equation}
    \frac{d}{d\chi}\left[(1-\chi^2)\frac{d\tilde{\vartheta}_0}{d\chi}\right]
    + c_{\log}
    =
    \frac{48(-1+15\chi^2-15\chi^4+\chi^6)}{(1+\chi^2)^5}.
\end{equation}
The general solution is
\begin{equation}
    \tilde{\vartheta}_{0}(\chi) = c_0 + \frac{1}{2}(-4 + c_{\log} + b_0) \log(1 - \chi) + \frac{1}{2}(-4 + c_{\log} - b_0) \log(1 + \chi) + 2 \log(1 + \chi^2) - \frac{4(\chi^4 + 4\chi^2 - 1)}{(\chi^2 + 1)^3}, 
\end{equation}
where $c_0, b_0$ are constants ($c_0$ can be removed by shift-symmetry). 

Choosing different values of the integration constants leads to solutions of different branches, whose physical interpretations are discussed in Sec.~\ref{sec:sGB}. 
Here we continue with the regular-pole branch, $c_{\log}=4$ and $b_0=0$.
We can also solve for the $n = 1$ solutions,
\begin{align}
    \bar{\vartheta}_{1}(\chi) &= 0, \\
    \tilde{\vartheta}_{1}(\chi) &= \frac{8\chi^2(\chi^4 + 4\chi^2 + 15)}{(\chi^2 + 1)^4} - 12 \chi \arctan \chi - 12.
\end{align}
The integration constants are fixed by imposing regularity at the pole and reflection symmetry. We can check that this matches the solution computed in Appendix~\ref{sec:sGB_Scalar_Field} using Legendre mode decomposition order by order in $\rho$ and $P_l(\chi)$, thus the near-horizon solution has the correct boundary conditions.

With the scalar field solution in hand, we can compute the metric corrections. Since the scalar field diverges logarithmically at the horizon, we cannot assume the horizon is regular (see Sec.~\ref{sec:sGB} and Fig.~\ref{fig:Scalar_Field_Horizon}). 
Instead, we expand the metric functions with a leading logarithmic term: 
\begin{equation}
    H_i(r, \chi) = j_i(\chi) \log(\rho) + h_i(\chi) + k_i(\chi) \rho \log \rho + p_i(\chi) \rho + o(\rho). 
\end{equation}
Furthermore, we must modify $\Xi$ in Eq.~\eqref{eq:metric} to be
\begin{equation}
    \Xi(r,\chi) = 1 + \zeta\left[\frac{\Upsilon(\chi) \log\rho}{r^4} + \frac{\Lambda}{r^4}\right], 
\end{equation}
where $\Upsilon(\chi)$ is a function to be determined. 
Taking this to compute the field equations, we find that $j_2(\chi)$, $h_2(\chi)$ are constants and $k_2(\chi) = 0$. 
Also, $j_1(\chi) = j_3(\chi)$.
Then, we can directly solve the remaining equations and yield\footnote{It is tempting to set $p_2(\chi) = 8/(1 + \chi^2) + \text{constant}$, since it makes all $j_i(\chi)$ and $\Upsilon(\chi)$ zero. However, it is unclear whether this profile of $p_2(\chi)$ is compatible with the higher-order equations, or the asymptotic behavior at spatial infinity. }
\begin{equation}\label{eq:sGB_j_i}
\begin{split}
    j_1(\chi) &= \frac{32\chi^2(1 - \chi^2)}{(1 + \chi^2)^3} + \frac{2\chi(1 - \chi^2) p'_2(\chi)}{1 + \chi^2}, \\
    j_2(\chi) &= 0, \\
    j_3(\chi) &= j_1(\chi), \\
    j_4(\chi) &= -\frac{32(2\chi^4 - 3\chi^2 + 1)}{(1 + \chi^2)^3} - 2(1 - \chi^2) p''_2(\chi) - \frac{2\chi(1 - \chi^2)p'_2(\chi)}{1 + \chi^2}, \\
    \Upsilon(\chi) &= \frac{32(2\chi^4 - 5\chi^2 + 1)}{(1 + \chi^2)^3} + 2(1 - \chi^2) p''_2(\chi) - 2\chi p'_2(\chi).
\end{split}
\end{equation}
The integration constants are eliminated by imposing reflection symmetry, regularity at the poles, and the absence of conical singularities $j_3(\chi = \pm1) = j_4(\chi = \pm 1)$. 
The function $p_2(\chi)$ parametrizes data not fixed at this order; importantly, the value of $\Lambda$ found below is independent of this freedom.
Next, we compute $h_i(\chi)$. Since $h_2(\chi)$ is a constant, we only need to find $h_{1,3,4}(\chi)$. The procedure is simple. First, at next-to-leading order, the $r\chi$-component of the field equations requires 
    \begin{equation}\label{eq:sGB_Extremal_BC}
        h_1(\chi) - h_3(\chi) = H_c + \frac{32}{1 + \chi^2} - 8\log(1 + \chi^2) - 2p_2(\chi).
    \end{equation}
Then, the next-to-leading order part of the $rr$- and $\chi\chi$-components of the field equations become a set of second-order differential equations for $h_3(\chi)$ and $h_4(\chi)$. Surprisingly, the $p_2(\chi)$ dependence drops out completely. Then, the equations can be directly solved. The integration constants are eliminated by enforcing regularity at the poles, and we find 
    \begin{equation}
        \Lambda = -\frac{1097}{16\sqrt{2}}.
    \end{equation}
    
For completeness, the full solution is 
\begin{equation}\label{eq:sGB_h_i}
\begin{split}
    h_1(\chi) &= c_1 + 2 h_2 - 2 p_2(\chi) - \frac{3291}{16\sqrt{2}} - \frac{1097\chi \sqrt{1 - \chi^2}\,Q(\chi)}{16\sqrt{2}(1 + \chi^2)} - \frac{22\chi \arctan\chi}{1 + \chi^2} \\
    &+ \frac{71744 + 634479\chi^2 + 727105\chi^4 + 915046\chi^6 + 434818\chi^8 + 38187\chi^{10} - 20739\chi^{12}}{1680(1 + \chi^2)^6}, \\
    h_2(\chi) &= h_2, \\
    h_3(\chi) &= h_1(\chi) + 2p_2(\chi) - H_c - \frac{32}{1 + \chi^2} + 8\log(1 + \chi^2), \\
    h_4(\chi) &= c_1 - \frac{1097}{8\sqrt{2}} + \frac{1097\chi Q(\chi)}{8\sqrt{2}\sqrt{1 - \chi^2}(1 + \chi^2)} + \frac{22\chi \arctan\chi}{1 + \chi^2} \\
    &+ \frac{25792 - 91263\chi^2 + 272080\chi^4 + 163530\chi^6 + 163048\chi^8 + 105261\chi^{10} + 26880\chi^{12}}{840(1 + \chi^2)^6}, 
\end{split}
\end{equation}
where 
\begin{equation}
    Q(\chi) = \arcsin\chi - \arctan\left(\frac{\sqrt{2}\chi}{\sqrt{1 - \chi^2}}\right).
\end{equation}
Several integration constants are fixed by imposing reflection symmetry and regularity at the poles. Additionally, one should also fix $h_3(\chi = \pm 1) = h_4(\chi = \pm 1)$ to remove conical singularities, which forces
\begin{equation}
    H_c = 2h_2 - \frac{3844}{105} - \frac{11\pi}{2} + 8\log{2}.
\end{equation}
Meanwhile, $c_1$ is another constant that should be matched to the exterior solution.

\section{Extremal Scalar Field Solution in sGB Gravity}\label{sec:sGB_Scalar_Field}

Instead of performing a near-horizon expansion, we can solve the extremal sGB scalar equation by decomposing the field into Legendre modes \cite{McNees:2015srl,Berti:2018cxi},
\begin{equation}
    \vartheta(r, \chi) = \sum_{l = 0}^{\infty} \vartheta_{l}(r) P_{l}(\chi)\,.
\end{equation}
Since $P_l(\chi)$ are regular at $\chi = \pm1$, this solution is guaranteed to have regular poles. 
Substituting into the scalar equation gives
\begin{equation}
    \frac{d}{dr} \left[(r - 1)^2 \frac{d \vartheta_{l}}{dr}\right] - l(l + 1) \vartheta_{l} = s_{l}(r), 
\end{equation}
where \cite{Lam:2025fzi}
\begin{equation}\label{eq:s_ell}
\begin{split}
    s_{l}(r) &= -\frac{2l + 1}{2} \int_{-1}^{+1} \frac{48 (r^6 - 15 r^4 \chi^2 + 15 r^2 \chi^4 - \chi^6)}{\Sigma^5} P_{l}(\chi) \,d\chi \\
    &= 4 {\rm Re} \, \bigg\{ \frac{(-1)^{\frac{l}{2}}}{r^{l+4}} \frac{\Gamma(\frac{1}{2})\Gamma(l+4)}{2^{l}\Gamma(l+\frac{1}{2})} \bigg[3\,{}_2F_1\left(\frac{l+4}{2},\frac{l+5}{2};l+\frac{3}{2};-\frac{1}{r^2}\right) - (l+5)\,{}_2F_1\left(\frac{l+4}{2},\frac{l+7}{2};l+\frac{3}{2};-\frac{1}{r^2}\right)\bigg]\bigg\}.
\end{split}
\end{equation}
The homogeneous solution is $A_{l} (r -1)^{l} + B_{l} (r - 1)^{-l-1}$, where $A_{l}$ and $B_{l}$ are integration constants. By variation of parameters, the ODE can be solved mode by mode exactly, and the particular solution is 
\begin{equation}\label{eq:vartheta_particular}
    \vartheta_{l}^{\rm p}(r) = \frac{1}{2l + 1} \left[(r - 1)^{l} \int_{\infty}^{r} \frac{s_{l}(x)}{(x - 1)^{l+1}} \,dx - \frac{1}{(r - 1)^{l+1}} \int_{1}^{r} s_{l}(x) (x - 1)^{l} \,dx\right]
\end{equation}
The most important one is the $l = 0$ mode. Given that 
\begin{equation}
    s_{l = 0}(r) = -\frac{16(3r^4 - 8r^2 + 1)}{(r^2 + 1)^4}, 
\end{equation}
the full solution is 
\begin{equation}
    \vartheta_{l = 0}(r) = 4 \log \left(\frac{r - 1}{\sqrt{r^2 + 1}}\right) - 2 \arctan{r} + \frac{2 (r+1) \left(r^2+r+2\right)}{\left(r^2+1\right)^2} + \pi.
\end{equation}
Thus, within the regular-pole, asymptotically flat exterior branch, the leading logarithmic divergence at the horizon cannot be removed by changing the homogeneous solution.
For $l > 0$, the integration constants $A_{l}$ and $B_{l}$ are fixed by regularity at the horizon and requiring $\vartheta_{l}(r) \sim r^{-l-1}$ at spatial infinity. 
For example, the first few Legendre modes are
\begin{align}
    \vartheta_{l = 2}(r) &= -24 (r-1)^2 \log \left(\frac{r-1}{\sqrt{r^2+1}}\right) + \frac{2\pi}{(r-1)^3} - \frac{\left(39 r^5-195 r^4+390 r^3-390 r^2+195 r-31\right) \arccot(r)}{(r-1)^3} \nonumber \\
    &+ \frac{15 r^7-84 r^6+195 r^5-288 r^4+325 r^3-284 r^2+145 r-40}{(r-1)^2 \left(r^2+1\right)^2}, \\
    \vartheta_{l = 4}(r) &= 60 (r-1)^4 \log \left(\frac{r-1}{\sqrt{r^2+1}}\right) - \frac{12\pi}{(r-1)^5} \nonumber \\
    &+\frac{3 \left(305 r^9-2745 r^8+10980 r^7-24780 r^6+34146 r^5-29610 r^4+16380 r^3-5940 r^2+1485 r-157\right) \arccot(r)}{4 (r-1)^5} \nonumber \\
    &- \frac{1}{4 (r-1)^4 \left(r^2+1\right)^2} \big[675 r^{11}-5640 r^{10}+21785 r^9-50200 r^8+81258 r^7 \nonumber \\
    &\qquad\qquad\qquad -102392 r^6+101358 r^5-77128 r^4+45299 r^3-19680 r^2+4665 r-384\big], 
\end{align}
and the odd $l$ modes vanish due to parity. 

\section{Norm of the Tidal Tensor}\label{sec:R2}

\begin{figure}[h!]
    \centering
    \includegraphics[width=0.5\linewidth]{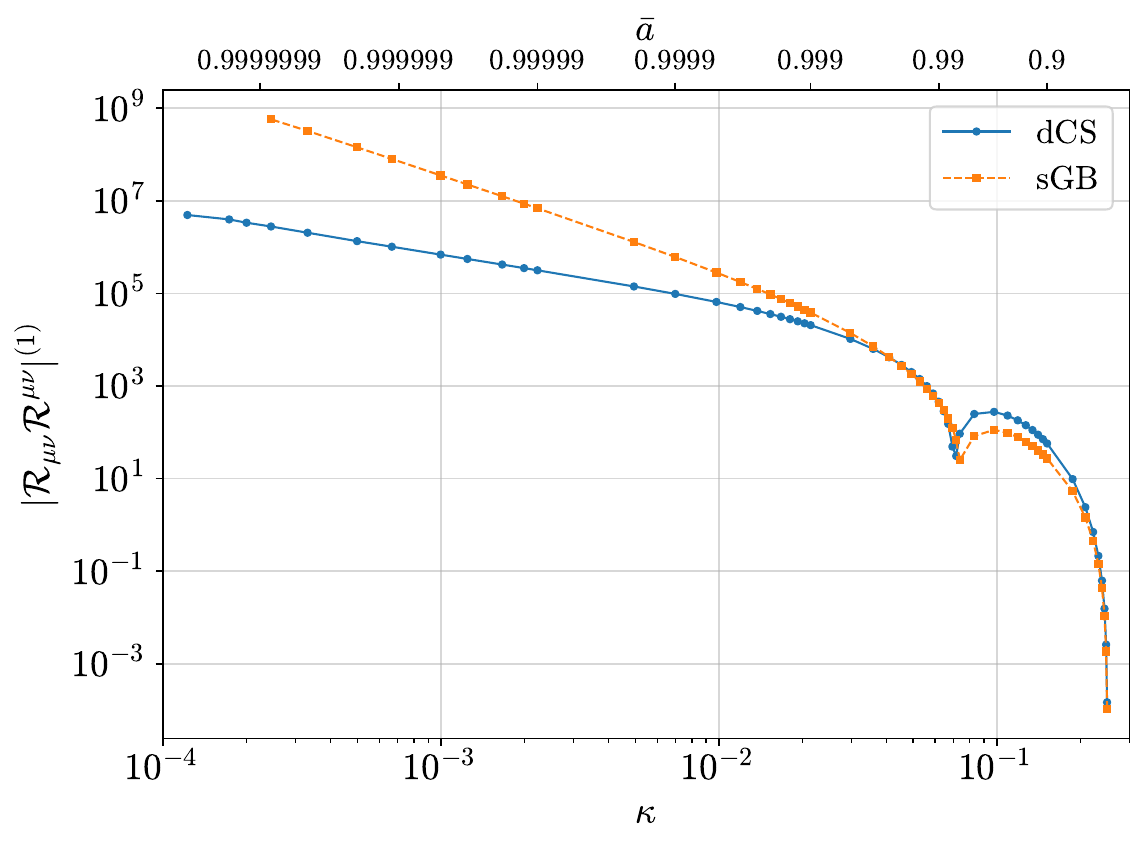}
    \caption{The correction of the square of the tidal tensor $|({\cal R}_{\mu\nu}{\cal R}^{\mu\nu})^{(1)}|$ against $\kappa$. }
    \label{fig:R2}
\end{figure}

In Sec.~\ref{sec:BreakdownEFT}, we have computed the tidal tensor assuming geodesics are separated in the $\chi$ direction.
In Fig.~\ref{fig:R2}, we plot the leading-order EFT correction
$|({\cal R}_{\mu\nu}{\cal R}^{\mu\nu})^{(1)}|$.
For the same ingoing null congruence and normalization used in Sec.~\ref{sec:BreakdownEFT}, this scalar is independent of the choice of separation vector.
In dCS and sGB gravity, fitting over the range $\kappa < 0.01$, it diverges as $\kappa^{-0.99}$ and $\kappa^{-2.07}$ respectively, similar to ${\cal R}_{\hat\chi\hat\chi}$. 

\twocolumngrid

\bibliography{ref}

\end{document}